\documentclass[manuscript]{aastex62}

\usepackage{amsmath,amsfonts,amssymb}
\usepackage{graphicx}
\usepackage{multirow}
\usepackage{tablefootnote}
\usepackage{threeparttable}
\submitjournal{AJ}

\shorttitle{
ALMA band-to-band phase referencing}
\shortauthors{Asaki et al.}
\begin{document}

\title{
  ALMA Band-to-band Phase Referencing: Imaging Capabilities on Long 
  Baselines and High Frequencies
}

\correspondingauthor{Yoshiharu Asaki}
\email{Yoshiharu.Asaki@alma.cl}

\author[0000-0002-0976-4010]{Yoshiharu Asaki}
\affil{Joint ALMA Observatory, 
        Alonso de C\'{o}rdova 3107, Vitacura, Santiago, 763 0355, Chile}
\affil{National Astronomical Observatory of Japan, \\
        Alonso de C\'{o}rdova 3788, Office 61B, Vitacura, Santiago, Chile}
\affil{Department of Astronomical Science, School of Physical Sciences, \\
        The Graduate University for Advanced Studies (SOKENDAI), 
        2-21-1 Osawa, Mitaka, Tokyo 181-8588, Japan}
\nocollaboration        

\author{Luke T. Maud}
\affil{ESO Headquarters, 
        Karl-Schwarzchild-Str. 2, D-85748 Garching, Germany}
\affil{Allegro, Leiden Observatory, Leiden University, 
        P.O. Box 9513, 2300 RA Leiden, The Netherlands} 
\nocollaboration

\author{Edward B. Fomalont}
\affil{Joint ALMA Observatory, 
        Alonso de C\'{o}rdova 3107, Vitacura, Santiago, 763 0355, Chile}
\affil{National Radio Astronomy Observatory, 
        520 Edgemont Rd., Charlottesville, VA 22903, USA}
\nocollaboration

\author{William R. F. Dent}
\affil{Joint ALMA Observatory, 
        Alonso de C\'{o}rdova 3107, Vitacura, Santiago, 763 0355, Chile}
\nocollaboration

\author{Loreto Barcos-Mu\~{n}oz}
\affil{National Radio Astronomy Observatory, 
        520 Edgemont Rd., Charlottesville, VA 22903, USA}
\affil{Joint ALMA Observatory, 
        Alonso de C\'{o}rdova 3107, Vitacura, Santiago, 763 0355, Chile}
\nocollaboration

\author{Neil M. Phillips}
\affil{ESO Headquarters, 
        Karl-Schwarzchild-Str. 2, D-85748 Garching, Germany}
\nocollaboration

\author{Akihiko Hirota}
\affil{Joint ALMA Observatory, 
        Alonso de C\'{o}rdova 3107, Vitacura, Santiago, 763 0355, Chile}
\affil{National Astronomical Observatory of Japan, \\
        Alonso de C\'{o}rdova 3788, Office 61B, Vitacura, Santiago, Chile}
\nocollaboration

\author{Satoko Takahashi}
\affil{Joint ALMA Observatory, 
        Alonso de C\'{o}rdova 3107, Vitacura, Santiago, 763 0355, Chile}
\affil{National Astronomical Observatory of Japan, \\
        Alonso de C\'{o}rdova 3788, Office 61B, Vitacura, Santiago, Chile}
\affil{Department of Astronomical Science, School of Physical Sciences, \\
        The Graduate University for Advanced Studies (SOKENDAI), 
        2-21-1 Osawa, Mitaka, Tokyo 181-8588, Japan}
\nocollaboration

\author{Stuartt Corder}
\affil{Joint ALMA Observatory, 
        Alonso de C\'{o}rdova 3107, Vitacura, Santiago, 763 0355, Chile}
\nocollaboration

\author{John M. Carpenter}
\affil{Joint ALMA Observatory, 
        Alonso de C\'{o}rdova 3107, Vitacura, Santiago, 763 0355, Chile}
\nocollaboration

\author{Eric Villard}
\affil{Joint ALMA Observatory, 
        Alonso de C\'{o}rdova 3107, Vitacura, Santiago, 763 0355, Chile}
\nocollaboration



\begin{abstract}
High-frequency long-baseline experiments with the Atacama Large 
Millimeter/submillimeter Array (ALMA) were organized to test the high  
angular resolution imaging capabilities in the submillimeter (submm) wave 
regime using baselines up to 16~km. Four experiments were conducted, 
two Band~7 (289~GHz) and two Band~8 (405~GHz) observations. 
Phase correction using band-to-band (B2B) phase referencing was used  
with a phase calibrator only $0^{\circ}.7$ away observed in 
Band~3 (96~GHz) and Band~4 (135~GHz), respectively.  
In Band~8, we achieved the highest resolution of $14 \times 11$~mas. 
We compared the synthesis images of the target quasar using 
20 and 60~s switching cycle times in the phase referencing. In Band~7, 
the atmosphere had good stability in phase rms 
($<0.5$~rad over 2~minutes) and there was little difference in image coherence 
between the 20 and 60~s switching cycle times. One Band~8 experiment 
was conducted under a worse phase RMS condition 
($>1$~rad over 2~minutes), 
which led to a significantly reduced coherence when using the 60~s 
switching cycle time. One of our four experiments indicates that the residual 
phase RMS error after phase referencing can be reduced to 0.16 rad 
at 289~GHz in using the 20~s switching cycle time. 
Such conditions would meet the phase correction requirement of image 
coherence of $>$ 70\% in Band~10, 
assuming a similar phase calibrator separation angle, emphasizing the need 
for such B2B phase referencing observing at high frequencies.
\end{abstract}

\keywords{instrumentation: high angular resolutions -- instrumentation: interferometers -- submillimeter: general}

%%%%%%%%%%%%%%%%%%%%%%%%%%%%%%%%%%%%%%
\section{
  Introduction
}\label{sec:01}
%%%%%%%%%%%%%%%%%%%%%%%%%%%%%%%%%%%%%%

The Atacama Large Millimeter/submillimeter Array (ALMA) is a very powerful 
instrument to investigate emissions in millimeter/submillimeter 
waves with very high angular resolutions 
%%%% REFERENCE
\citep{Bachiller2008}. 
With the 16~km ALMA configuration, the spatial resolutions will be 
12, 7, and 5~milliarcseconds (mas) at the observing frequencies of 
400, 650, and 850~GHz, respectively. 
With these high angular resolutions, ALMA is expected 
to reveal unseen astrophysical properties in a variety of 
celestial objects in submillimeter waves. 

The long-baseline capabilities of ALMA have been tested since 
ALMA early science operation started in 2011 
%% REFERENCE
\citep{Matsushita2012, 
%% REFERENCE
Matsushita2016,
%% REFERENCE
Asaki2016}. 
ALMA long-baseline campaigns (LBCs) were organized in 2014 
and 2015 to accomplish the most extended array configuration with 
16~km baselines in order to make science verification observations in 
Bands~3, 6, and 7 
%% REFERENCE
\citep{ALMApartnership2015a,
%% REFERENCE
ALMApartnership2015b,
%% REFERENCE
ALMApartnership2015c,
%% REFERENCE
ALMApartnership2015d}. 
ALMA has opened 16~km baseline observations to the user
community up to Band~7 frequencies in Cycle~7 
($\geq 0.8$~mm wavelength, or $\leq 373$~GHz), 
with the ability to regularly achieve resolutions higher than $\sim 20$~mas. 
For example, ALMA imaged scenes of star formation in 
a distant gravitationally lensed galaxy SDP.81 at $z=3.042$ 
with the angular resolution of $\sim 23$~mas in Band~7, 
corresponding to a spatial scale of $\sim 180$~pc 
%%REFERENCE
\citep{ALMApartnership2015c}. 
ALMA allows us to address planet formation around very young 
solar-type stars in the Milky Way through observations of protoplanetary 
disks, where the dust temperature distribution, the nature of dust particles, 
and the gas dynamics can be investigated in unprecedented detail.
For example, 
ALMA revealed a protoplanetary dust disk with a pattern of bright rings and dark gaps 
surrounding HL~Tau with the angular resolution of 25~mas 
%% REFERENCE
\citep{ALMApartnership2015b} 
and 
%% REFERENCE
TW~Hya with the angular resolution of 20~mas in Band~7 
\citep{Andrews2016}, 
corresponding to 3.5 and 1~au, respectively, at the target distance. 
It is also expected to image a few au scale circumplanetary disks 
around massive planets in such protoplanetary disks 
%% REFERENCE
\citep[e.g.][]{Isella2019, 
%% REFERNECE
Tsukagoshi2019}. 
Recently the magnetic field structure around a very young protostar, 
MMS~6, in the Orion Molecular Cloud-3 
was studied using dust emission polarization with the angular resolution 
of $\sim 25$~mas, corresponding to $\sim 10$~au 
%% REFERECE
\citep{Takahashi2019}. 
For evolved stars, ALMA captured a bright spot on the asymmetric photosphere 
of Betelgeuse with the 13~mas angular resolution, corresponding to 2~au 
%% REFERENCE
\citep{OGorman2015B}, 
and detected the circumstellar mass-loss gas with the angular resolution of 18~mas 
in Band~7 
%% REFERENCE
\citep{Kervella2018}. 
ALMA is also an excellent tool to study solar system objects: 
an asteroid, 3~Juno, was observed with the angular resolution of 42~mas in Band~6, 
corresponding to 60~km at 1.97~au 
and was able to resolve thermal emission from the surface 
of the main belt asteroid, to measure the asteroid geometric shape, rotational period, 
and soil characteristics 
%% REFERENCE
\citep{ALMApartnership2015d}. 

The highest angular resolutions in radio interferometry can be realized 
with the combination of the longest baseline possible and highest observing 
frequencies available. To achieve the highest angular resolutions with ALMA, 
interferometer phase stability for the 16~km baselines in high-frequency (HF) 
bands 
(Bands~8--10, or the frequency range of 385--950~GHz) is crucial. 
Atmospheric fluctuations increase the interferometer phase error 
as a function of baseline length 
%% REFERENCE
\citep{Matsushita2017}, 
so that the longer the baseline, the worse the phase stability, and thus 
the synthesized image becomes blurred 
%% REFERENCE
\citep{Carilli1999}. 
The interferometric phase error can usefully be translated into a coherence factor. 
%%%%
Assuming that the residual phase variations in the visibility data are pseudo-random, 
the coherence factor (percentage decrease of the peak intensity of the image) can 
be evaluated by 
$\exp{({-{\sigma_{\Phi}}^2/2})}$, where $\sigma_{\Phi}$ is the phase rms in rad 
%% REFERENCE
\citep{TMS2001}. 
This value is directly related to the image performance of radio interferometers. 
The phase RMS has to be decreased to 0.84~rad in order to achieve 
coherence values above 70\%. Since the interferometer phase is proportional to 
a product of an electrical path length error of the electromagnetic wave and 
observing frequency, the corresponding path length fluctuation is 100, 61, and 
47~$\mu$m~RMS at 400, 650, and 850 GHz, respectively. 
Reducing the atmospheric phase fluctuations is therefore a critical requirement for 
any calibration technique in order to achieve successful ALMA HF long-baseline 
observations.

To reduce the atmospheric phase fluctuations, phase referencing is 
a frequently adopted technique 
for interferometer phase correction 
%% REFERENCE
\citep{Beasley1995,
%% REFERENCE
Asaki2007}. 
The general implementation of phase referencing is to frequently 
observe a phase calibrator, typically a bright point-like quasar (QSO), 
close to the target source at the same frequency. 
This is referred to as in-band phase referencing and is the common 
phase correction method used for ALMA and other radio and millimeter 
interferometers. 
ALMA has been preparing 
a millimeter-wave bright QSO catalog (ALMA calibrator source catalogue\footnote{
/https://almascience.nrao.edu/alma-data/calibrator-catalogue}) 
for in-band phase referencing since the early science operations. 
During the ALMA mission concept stage and through 
various site tests, it was realized that calibration of the longer baselines with 
a standard scheme would require nearby calibrators within a few degrees, limited 
by both potential antenna slew speeds and intrinsic site phase stability. It was also 
predicted that a few thousand calibrators would be available on the sky at 
a wavelength of 3~mm and thus could easily be found close to any target 
%% REFERENCE
\citep[e.g.][]{Holdaway1992,
%% REFERENCE
Holdaway2001a}. 
However, the flux density 
%$F_{\mathrm{\nu}}$ 
of a typical QSO is proportional to 
$\nu^{-0.8}$ in the millimeter/submillimeter-wave regimes 
%% REFERENCE
\citep{Hardcastle2000} 
and the system 
noise (both from the sky and receivers) increases, meaning that the integration 
time to achieve sufficient signal-to-noise ratio (S/N) becomes too large. 
This combination makes it 
unlikely to find a bright enough nearby phase 
calibrator for a randomly located target  
%% REFERENCE
\citep{Asaki2020a}. 

In order to mitigate the difficulty in finding a phase calibrator at HF, 
the so-called band-to-band (B2B) phase referencing provides 
an alternative technique, employing observations of a 
sufficiently close phase calibrator at a low-frequency (LF) 
Band. The use of a phase calibrator closer to the target 
than would otherwise be possible with in-band phase correction at HF 
will provide more accurate phase referencing 
%% REFERENCE
\citep{Holdaway2004b}. 
This B2B phase referencing technique was firstly demonstrated using the 
Nobeyama Millimeter Array between 148 and 19.5~GHz 
%% REFERENCE
\citep{Asaki1998} 
and has since been used successfully at various other facilities. 
The Korean very long baseline interferometry (VLBI) network (KVN) can observe 
22, 43, 86, and 129~GHz simultaneously with 
a quasi-optical receiver system 
%% REFERENCE
\citep{Han2013} 
and achieves multiwave B2B phase referencing, which is called 
source frequency phase referencing. 
It has been demonstrated in VLBI observations of active galactic nuclei 
%% REFERENCE
\citep{Rioja2014}
and astrophysical maser sources 
%% REFERENCE
\citep{Dodson2014}.
A similar multifrequency phase correction has been successfully applied to 
Combined Array for Research in Millimeter-wave Astronomy (CARMA) 
observations at 227~GHz 
where a subset of antennas positioned close to the main array 
observed at 30~GHz and were used to 
correct the 227~GHz target phase using the 30~GHz delay 
measurements in the CARMA paired antenna calibration system 
%% REFERENCE
\citep[C-PACS;][]{Perez2010, 
%% REFERENCE
Zauderer2016}. 

B2B phase referencing is currently being implemented at ALMA. 
The motivation of the research presented in this paper is to verify 
the technical feasibility of B2B phase referencing and evaluate 
its performance in synthesized images with the high angular resolutions. 
In this paper, we report on HF long-baseline (14--16~km) 
image capability tests for a point source QSO J2228$-$0753 in 
Band~7 (289~GHz) and Band~8 (405~GHz) observed 
together with a closely located 
phase calibrator, J2229$-$0832, in Band~3 (96~GHz) and Band~4 (135~GHz), 
respectively, 
to demonstrate a quasi end-to-end ALMA science 
observation.  
The data were obtained during the ALMA HF LBC 
%% REFERENCE
\citep[HF-LBC-2017; ][]{Asaki2020a}. 
In 
%%%% SECTION
Section~\ref{sec:02}, 
the basic strategy of B2B phase referencing for HF long-baseline 
observations is mentioned. 
In 
%%%% SECTION
Sections~\ref{sec:03} 
and 
%%%% SECTION
\ref{sec:04},  
we outline our ALMA B2B phase referencing observation experiments and data reduction, respectively. 
%%%% SECTION
Section~\ref{sec:05} 
provides the results of the experiments, 
which are discussed in 
%% SECTION
Section~\ref{sec:06}. 
We summarize 
this feasibility study in 
%%%% SECTION
Section~\ref{sec:07}. 
A summary of the HF-LBC-2017, which these tests are part of, is presented by 
%% REFERENCE
\cite{Asaki2020a}, 
while further tests investigating calibrator separation angles and switching 
cycle times are presented by 
%% REFERENCE
\cite{Maud2020}.

%%%%%%%%%%%%%%
\section{
    Basic strategy of B2B phase referencing
}\label{sec:02}
%%%%%%%%%%%%%%

The main goal of the interferometer phase correction is 
to remove systematic phase errors due to instrumental electrical path length 
errors and a priori antenna 
position errors, atmospheric delay models over the array, and short-term 
atmospheric delay variations.  An important delay correction produced by 
water vapor over each 12~m antenna is obtained by water vapor radiometers 
%% REFERENCE
\citep[WVRs;][and references therein]{Nikolic2013} 
on each antenna. The water vapor emission in the direction of the radio source 
measured with the WVR is converted into delay determined every second. 
The WVR phase correction is always applied to the ALMA data 
and typically removes the majority of the water vapor delay 
%% REFERENCE
\citep{Matsushita2017, 
%% REFERENCE
Maud2017}, 
although it is somewhat dependent on the quantity of the water vapor content 
in the atmosphere as precipitable water vapor (PWV), the depth of the water 
in a column of the atmosphere to the zenith. 

In 2017, HF-LBC-2017 was organized to test the submillimeter-wave 
(wavelength of shorter than 1.1~mm) 
imaging capability of ALMA using the 
longest baselines up to 16~km 
%% REFERENCE
\citep{Asaki2020a}. 
One focus of the campaign was to extensively test the B2B phase referencing 
method and begin providing appropriate observing scripts and data reduction scripts.
In B2B phase referencing, a close phase calibrator is observed at an LF. 
The LF phase of the phase calibrator is multiplied by the HF/LF ratio. 
%% REFERENCE
\cite{Asaki2020a} 
describe the realization and implementation of B2B phase referencing in ALMA 
in detail. 

An important observation parameter in phase referencing (in-band or B2B) is 
the switching cycle time $t_{\mathrm{swt}}$ that is measured by the time 
difference between two calibrator scans. 
For ALMA observations with the longer baselines, the typical switching cycle time 
is 72~s. In many cases shorter switching cycle times provide more accurate phase 
referencing and are investigated.
Our experiments adopted a much faster switching cycle time of 20~s 
because HF observations may require more frequent visits 
to a phase calibrator to compensate for the atmospheric phase fluctuations. 
The ALMA antennas can quickly change their position by several degrees in a few seconds 
to accommodate such a quick source change, which is most important at HF, where 
the atmospheric fluctuations are most variable.

Use of B2B phase referencing requires an additional phase correction to remove 
the instrumental phase offset difference between the two respective Bands. 
This can be achieved through a cross-band calibration, 
which is referred to as differential gain calibration (DGC). In the DGC, 
a bright QSO is observed at both an HF and LF, using frequency switching, 
such that the phase offset difference can be solved. 
If the phase offset difference is time constant, a single DGC measurement is enough. 
One important question at the ALMA system level is whether the temporal instrumental 
phase stability is sustained in the case where the HF phase is corrected by the LF phase. 
This uncertainty 
is due to the fact that the highly stabilized local oscillator (LO) signal 
does not pass through identical signal paths in the instrumentation for the 
LF and HF data 
during B2B phase referencing. Intrinsically 
the HF phase to be corrected flows through the HF receiver, while the correcting phase 
at the LF passes though the LF receiver. Thus, it is important to confirm that the phase 
offset differences 
between the respective Bands are stable with time during B2B 
phase referencing as has been confirmed for continuous single (in-band) observations 
%% REFERENCE
\citep[e.g.][]{Matsushita2012}. 
In addition, the atmospheric phase variations add a pseudo-random phase difference 
between HF and LF in DGC; hence, the frequency switching should be as fast as 
possible.

%%%%%%%%%%%%%%%%%%%%%%%%%%%%%%%%%%%%%%%%%%%%%%%%
\section{
  Observations
}\label{sec:03}
%%%%%%%%%%%%%%%%%%%%%%%%%%%%%%%%%%%%%%%%%%%%%%%%

\begin{figure}[htbp]
\begin{center}
\begin{tabular}{c}
\includegraphics[width=70mm]{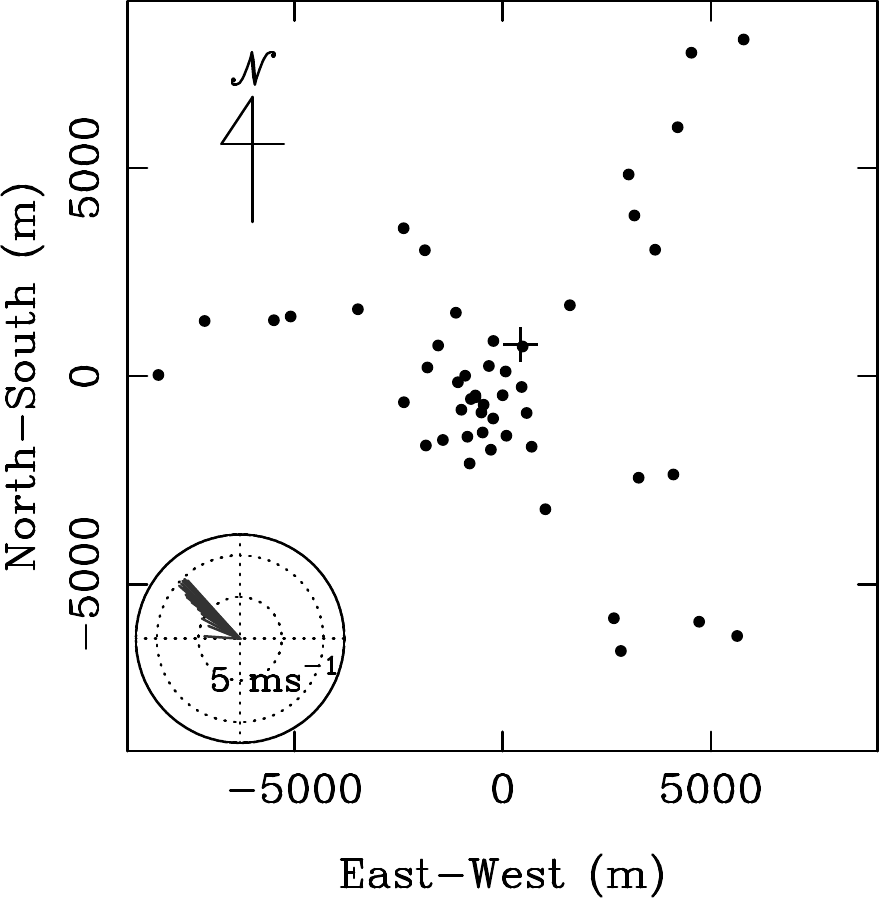}
\hspace{20mm}
\includegraphics[width=70mm]{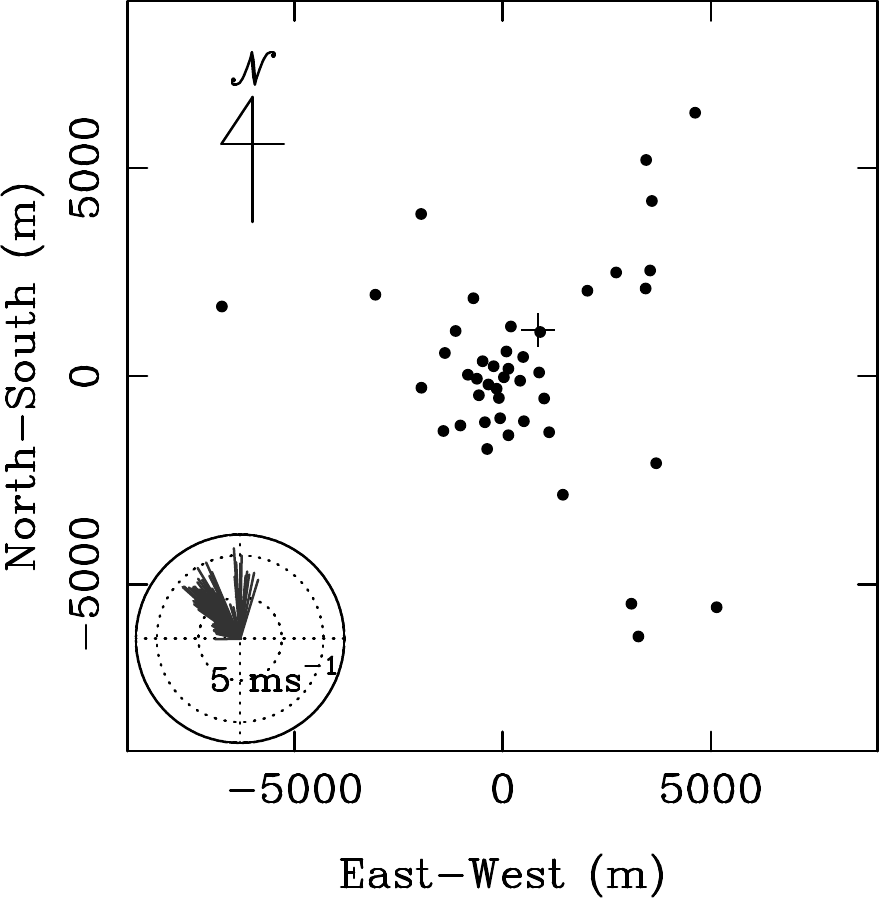}
\end{tabular}
\caption{
ALMA antenna
configuration of the experiments on 2017 October 9 (left)
and on 2017 November 2 (right).
The filled circles represent the 12~m antenna position.
The bottom left circle in each panel shows the wind speed and direction during the experiment
measured with
the weather station located at the cross point. 
The inner and outer circles represent the wind velocity of 5 and 10~m~s$^{-1}$, respectively. 
}
\label{fig:01}
\end{center}
\end{figure}

We conducted the observing experiments in 2017 October and November   
with the maximum baseline lengths of 16 and 14~km,  respectively, as shown in 
%% FIGURE
Figure~\ref{fig:01} 
with 40--50 12~m antennas.
Overall four experiments were conducted, two in October (9th and 10th) and 
two in November (2nd and 3rd). We arranged B2B phase referencing 
experiments at two specific frequency pairs. 
One is at 289~GHz in Band~7 for 
the target with the phase calibrator observed at 96~GHz 
in Band~3 (Band~7-3 experiment), 
while the other is at 405~GHz in Band~8 for the target with the 
phase calibrator 
at 135~GHz in Band~4 (Band~8-4 experiment). 
Each epoch consists of one Band~7-3 and one Band~8-4 experiment. 
For convenience, the first- and second-epoch Band~7-3 experiments are 
referred to as Band~7-3(1) and Band~7-3(2), respectively, while the Band~8-4 
experiments are denoted Band 8-4(1) and Band 8-4(2). 
Details are listed in 
%%%% TABLE
Table~\ref{tbl:01}.

The observed sources are listed in 
%%%% TABLE
Table~\ref{tbl:02}. 
We selected the QSO J2228$-$0753 as a target at the HF, while the 
QSO J2229$-$0832, with a separation angle of $0^{\circ}$.7,
was observed as a phase calibrator at the LF. 
For DGC, 
we selected a very bright QSO J2253$+$1608  
located $25^{\circ}$ away from the target. 
Note that the phase offset difference measured by a DGC source is 
independent of the QSO location in the sky, since it is predominately 
determined by internal electronic phase differences between the HF and LF 
%% REFERECE
\citep{Asaki2020a}.
We switch between the HF and LF using the same frequency of 
the photonic LO signal 
%% REFERENCE
\citep{Shillue2012}
in order to minimize overhead times for the frequency switching 
%% REFERENCE
\citep{Asaki2020a}. 
This allows us to change frequency with only a $\sim 2$~s delay. 

ALMA science operations have two criteria to assess go/no-go of a given 
user observation depending on weather conditions.
First is the amount of the water vapor (PWV) because of the absorption of the radio signal. 
A loss of 20\% absorption corresponds to a PWV of 1 and 0.7~mm at frequencies of 
350 and 450~GHz, respectively, and these are guidelines of whether observations should be 
started. The PWV is also associated with short-term phase variations over each antenna. 
The variations are somewhat correlated with the PWV value but also depend on the water vapor 
clumpiness and wind velocity.  Second is an estimate of the interferometer phase stability 
determined from the phase behavior in the previous observation, and/or a short observation 
known as go/no-go. 
The nominal criterion used from the above test is that the phase RMS with the time interval 
of 2~minutes for baselines about half the maximum length is less than $\sim 1$~rad.   
At the start of all our experiments, the two criteria were passed. 
However, note that the conditions can change during an observation and over the sky, 
so these go/no-go criteria are only approximate.

\begin{figure}[htbp]
\begin{center}
\begin{tabular}{l}
\includegraphics[height=55mm]{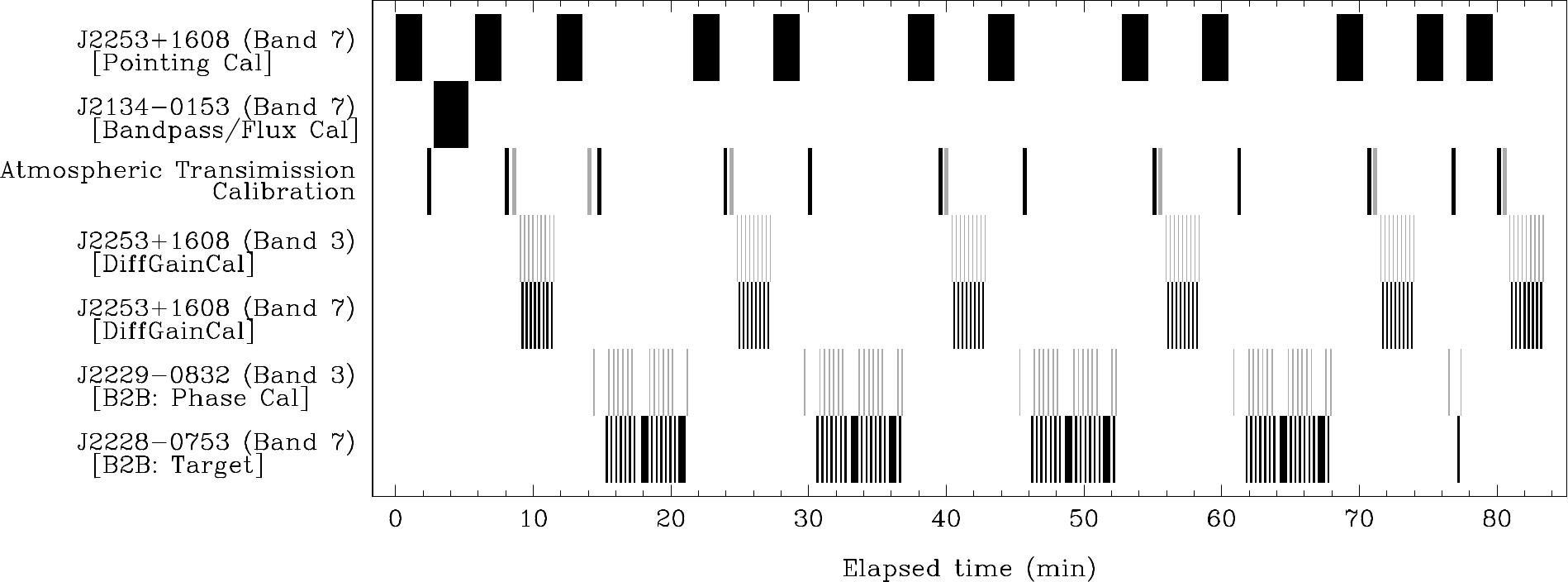}
\\
\includegraphics[height=55mm]{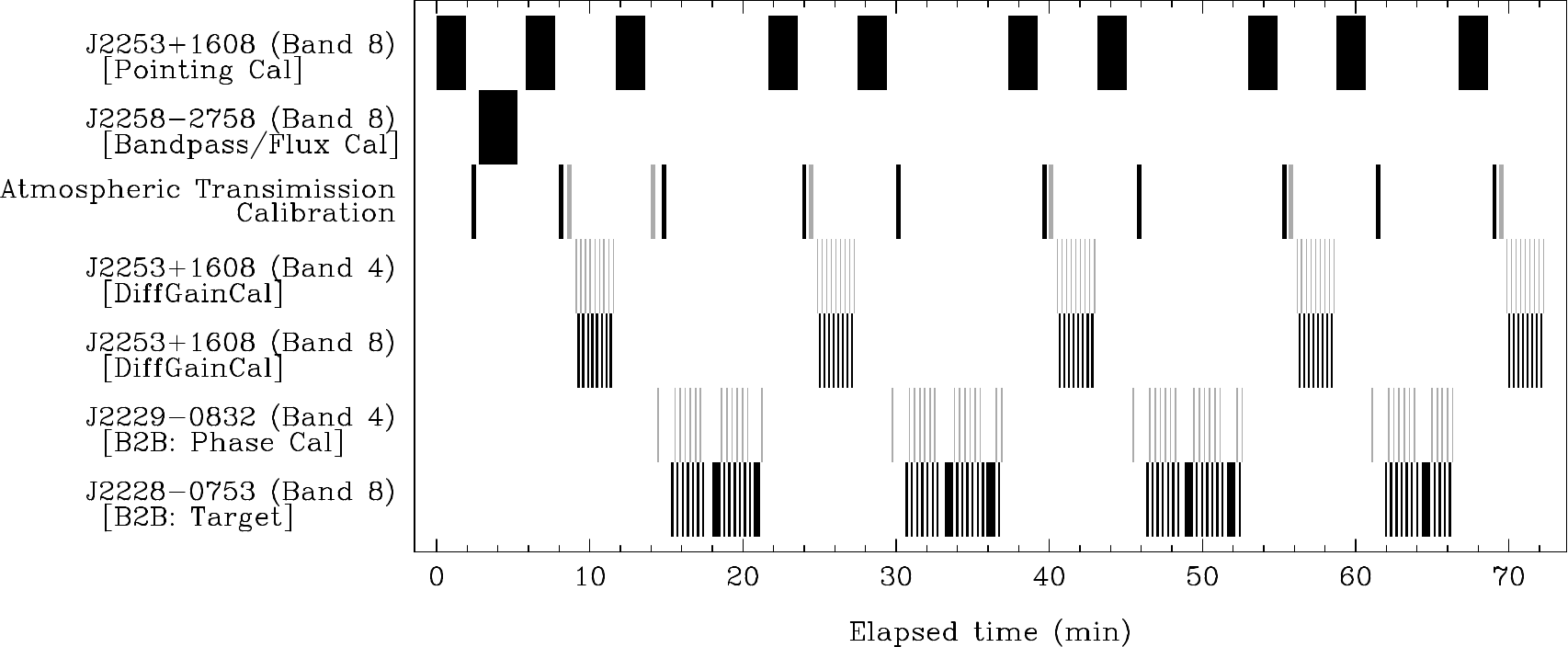}
\end{tabular}
\caption{
Observation schedule of the second-epoch experiments. 
Top: Band~7-3 experiment on 2017 November 2. 
Bottom: Band~8-4 experiment on 2017 November 3. 
The black and gray boxes represent HF and LF scans, respectively. 
J2134$-$0153 was observed as a flux calibrator, which was not used in the data reduction. 
}
\label{fig:02}
\end{center}
\end{figure}

In order to demonstrate general ALMA science observations, 
we prepared experimental Scheduling Blocks 
%% REFERENCE
\citep[SBs;][]{Nyman2010} 
containing all the observation information required for B2B phase referencing 
and DGC, as well as normal observing sequences such as system noise 
temperature measurements, pointing calibration, and flux calibration at the HF. 
A DGC block begins and ends with an LF scan and consists of 
eight HF and nine LF scans executed by turn. 
The experimental SBs for the second epoch are depicted in 
%% FIGURE
Figure~\ref{fig:02}. 
A single observation consists of repetitions of the 
B2B phase referencing block for the target and phase calibrator 
and regular visits to the DGC block where we only switch the frequency 
while pointing at 
the DGC source.  
The B2B phase referencing block between the target and phase calibrator 
lasts 12~minutes, after which a 2.5~minute DGC block 
is inserted. Such a frequent insertion of the DGC block was made not 
only to check the DGC solution repeatability and stability but also to avoid 
a heavy load to the antenna control computers by a long sequence of antenna 
pointing changes.

Each Band amplifies two linear polarizations
($X$ and $Y$) separately at each 12~m antenna. 
The amplified signal is split into four intermediate-frequency signals,
which are referred to as basebands (BBs) with the bandwidth of 2~GHz.
Two of the four BBs are transmitted from the upper side band (USB) of the ALMA 
front-end receiver, and the other two are transmitted from the lower side band (LSB). 
The four BBs with the $X$ and $Y$ polarizations are digitized at the antenna and 
transferred to the 12~m array correlator
%% REFERENCE
\citep{ALMA_TH_book_C8}. 
In our experiments, 
the digitized BBs were filtered out to have a 1.875~GHz bandwidth each 
with 128 frequency 
channels for two polarization pairs of $XX$ and $YY$ to form a spectral window 
(SPW). The center frequencies of the HF SPWs are 
282.5 (LSB), 284.4 (LBS), 294.5 (USB), and 296.5 (USB) GHz in Band~7, and 
398.0 (LSB), 399.9 (LSB), 410.0 (USB), and 412.0 (USB) GHz in Band~8. 
The integration time used for 
recording each of the 
four SPWs at both the HF and the LF respectively 
was 1.01~s.

For the B2B phase referencing blocks, a typical science target scan length was 
8~s at the HF, while the phase calibrator scan length at the LF was 6~s. 
For the DGC blocks, the above frequency switching sequence was adopted, 
and the scan length for the HF and LF DGC scan  
was typically 8 and 6~s, respectively. 
The antenna slew and/or the frequency switch 
are done simultaneously and take 2--3~s.
The total length for the one switching cycle is $\sim 20$~s in both the 
B2B phase referencing sequence and DGC blocks. 

Since the selected DGC source is one of the brightest sources in the sky, 
we also used DGC scans for a bandpass calibration at both the LF and HF, 
so a separate bandpass scan was not prepared in the SBs. 
We also used the HF DGC scans for the flux calibration as well in the data 
reduction, 
while flux calibration at the LF is not required, as we are only interested 
in the transfer of the phase solutions. During operations, it is envisaged that separate 
bandpass and flux calibrators may be observed.

%%%%%%%%%%%%%%%%%%%%%%%%%%%%%%%%%%%%%%%%%%%%%%%%%%%
\section{
  Data Reduction
}\label{sec:04}
%%%%%%%%%%%%%%%%%%%%%%%%%%%%%%%%%%%%%%%%%%%%%%%%%%%

Data reduction was carried out using the Common Astronomy Software Applications 
(CASA) software 
%% REFERENCE
\citep{McMullin2007}, 
and we have developed a semi-automatic CASA data reduction 
python script for our experiments. 
The offline WVR phase correction (WVRGCAL) 
%% REFERENCE
\citep{Nikolic2012} 
was applied for all the experiment data.  
The obtained antenna-based 
water vapor brightness is translated into an antenna-based 
phase solution. 
The obtained PWVs in the course of WVRGCAL are listed in 
%% TABLE
Table~\ref{tbl:03}. 
WVRGCAL could reduce the atmospheric phase fluctuations 
even in the low PWV conditions ($<1$~mm), 
with a reduction of the phase RMS between 4\% and 35\% 
on the longest baselines. 

Most of the target scan lengths enclosed by the phase calibrator scans 
have the switching cycle time of 20~s, while some of the 
target scans have longer switching cycles (typically,  $\sim 1$~minute) 
because the ALMA control software inserts a $\sim20$~s system noise temperature 
measurement just after the 
first phase calibrator scan in the target-phase calibrator block.
Since one of the purposes of our experiments is to see the 
phase correction efficiency with a short switching cycle time, 
the data reduction script flagged out the target scans with such 
a longer switching cycle time (at most 6--7~scans of the 
$\sim 100$ during the observations).  

There are two phase transfer processes in the data reduction: the 
first is for the DGC, in which J2253$+$1608's HF scan phase is 
corrected using its LF scan phase. The second is for the target J2228$-$0753 
whose phase is corrected using the phase calibrator J2229$-$0832. 
The DGC phase transfer is performed prior to the target--phase 
calibrator process. In the DGC, we first obtain the SPW-dependent 
LF phase offset by averaging the LF DGC scans in time for each SPW. 
The obtained LF phase offset is self-applied to the LF scans of the DGC source, 
as well as to the phase calibrator. After removing the LF phase offset from the 
LF scans, the LF DGC SPWs are averaged for phase-up, and the phase transfer 
is conducted to remove the atmospheric phase fluctuations from the 
HF DGC scans using LF DGC scans temporally closest. 
The HF DGC phase can be then averaged in time for each SPW at each 
DGC block to obtain the SPW-dependent HF phase offset. 
In this paper, the HF DGC phase offset is referred to as the DGC solution. 
One DGC solution is obtained by averaging the eight HF DGC scans 
from a DGC block after the phase transfer. To carry out the phase transfer 
from the phase calibrator to the target, the SPWs of the LF phase calibrator 
are averaged for the phase-up and then applied to the HF target phase for 
each SPW together with the aforementioned DGC solution. 
One of the most essential parts in B2B phase referencing is the 
temporal stability of the DGC solution, which is described in 
%% SECTION
Section~\ref{sec:05-02}. 
In the course of the WVR phase correction, antenna position calibration, 
bandpass calibration, flux calibration, and B2B phase referencing, 
significant outliers in either of the phase or amplitude were flagged out from 
the visibility data. 

The visibility data of all four HF SPWs and both polarization pairs are combined 
using multifrequency synthesis imaging in the CLEAN algorithm using a fixed value of 
50~CLEANing iterations. We used Briggs weighting with the robustness 
parameter of 0.5 in imaging and the pixel size of 2~mas for a $512 \times 512$ 
square mas area ($256 \times 256$ pixels). In the CLEAN deconvolution, 
we set a 15-pixel-radius circle mask at the phase tracking center. 

We additionally made phase self-calibration 
%% REFERENCE
\citep{Schwab1980} 
with the solution 
interval of the target on-source scan length of 8~s 
for the purpose of investigating the effectiveness of B2B phase referencing: 
the self-calibrated image provides the synthesized image of the target 
without phase errors on timescales longer than 8~s, 
so that comparison of the images with and without the phase self-calibration 
can inform us of how much B2B phase referencing can be effectively applied 
for the phase correction. 
In general, if a target source is bright enough, self-calibration effectively works 
to correct the interferometer phase without frequent phase calibrator scans 
%% REFERENCE
\citep{Brogan2018, 
%% REFERENCE
Cornwell1981}. 
However, it is not plausible to assume that high S/N is guaranteed 
for HF observations in general considering the notably higher system noise 
temperatures and lower antenna efficiencies when compared to the 
lower-frequency Bands. In addition, 
target sources observed with the higher angular resolutions are likely to be resolved. 
Therefore, S/N on the longest baselines could be further reduced, and it may be more 
difficult to perform self-calibration compared with the point-source case.
This means that self-calibration cannot be a default phase correction plan to improve the 
image quality of the target, so the effectiveness 
of the self-calibration for ALMA HF long-baseline observations is out of scope 
in this paper.

%%%%%%%%%%%%%
\section{
  Results
}\label{sec:05}
%%%%%%%%%%%%%
\subsection{
  Reassessment of Weather Conditions 
}\label{sec:05-01}
%%%%%%%%%%%%%

The PWV and phase stability were reassessed from the data to confirm the 
go/no-go judgment of the experiments. The PWV was obtained by WVRGCAL. 
The phase stability was estimated from the phase RMS of the HF DGC scans 
in each of the single DGC blocks (time interval $\sim 2$~minutes each). 
The HF phase data after WVRGCAL but before phase transfer were used 
to obtain the phase stability for the baselines between an antenna located 
at the almost array center (reference antenna) and the outermost antennas 
(the upper quartile of the total number 
of the antennas). This is referred to as the 2-minute phase stability in this paper, 
which is equivalent to the phase stability in the go/no-go. 
The reassessed weather conditions are listed in 
%% TABLE
Table~\ref{tbl:03} 
along with other weather data (an average wind speed and wind direction).  
Note that the 2-minute phase stability is listed with the minimum and maximum 
values during each experiment. 
During the Band~7-3 experiments, the weather conditions were completely 
satisfied. On the other hand, Band~8-4(1) has rather higher PWV ($>$0.7~mm), 
although this overall just means a higher image noise for the observing time than 
is generally achieved. 
At Band~8-4(2), the 2-minute phase stability is satisfied in the beginning and end of 
the experiment, while it increased to almost double the upper limit in the middle, 
indicating that the atmospheric phase stability was somewhat marginal during the 
Band~8-4(2) experiment. 

%%%%%%%%%%%%%
\subsection{
  Temporal stability of DGC solutions
}\label{sec:05-02}
%%%%%%%%%%%%%

\begin{figure}[htbp]
\begin{center}
\begin{tabular}{c}
\includegraphics[height=130mm]{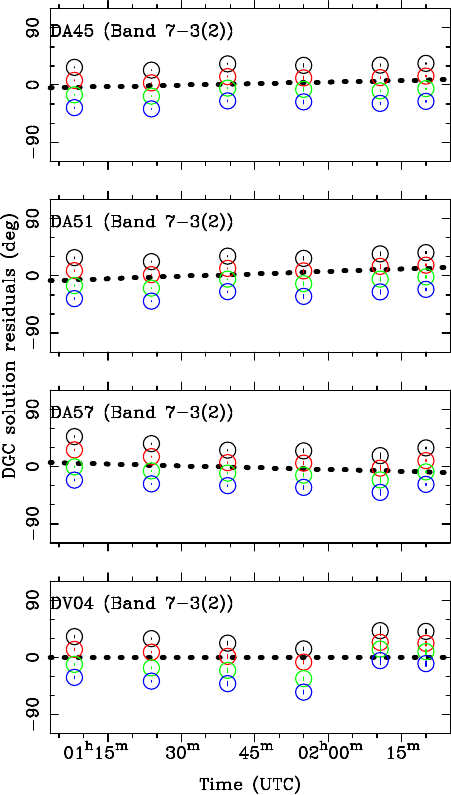}
\hspace{10mm}
\includegraphics[height=130mm]{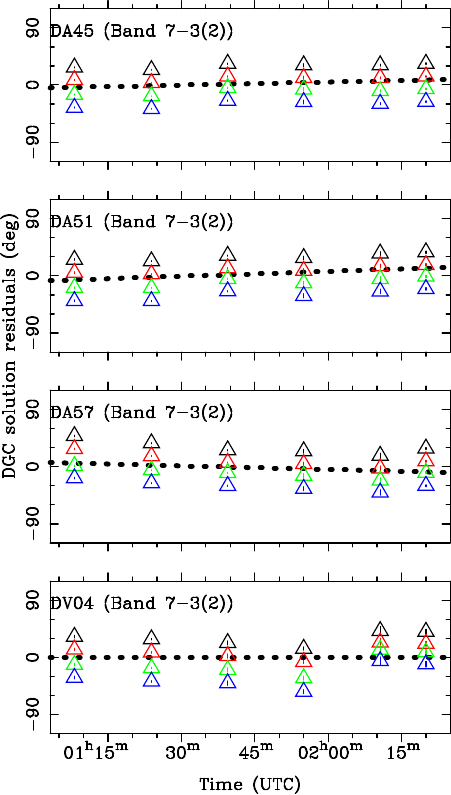}
\end{tabular}
\caption{
Antenna-based 
DGC solutions of J2253$+$1608 after 
adding an arbitrary 
phase offset to each SPW at Band~7-3(2) for plotting purposes. 
The horizontal and vertical axes are observation time and 
residual of the DGC phase solutions 
in degrees.   
The left and right panels show the polarization pair of $XX$ and $YY$ with the open circles and 
triangles, respectively. The color differentiates four SPWs. 
The dotted lines represent the linear trend of the DGC solutions for each antenna (see 
%% SECTION
Section~\ref{sec:05-02}).
The top two panels show antennas with small 
DGC solution RMS, 
while the bottom two show the worst antennas. 
}
\label{fig:03}
\end{center}
\end{figure}

\begin{figure}[htbp]
\begin{center}
\begin{tabular}{c}
\includegraphics[height=130mm]{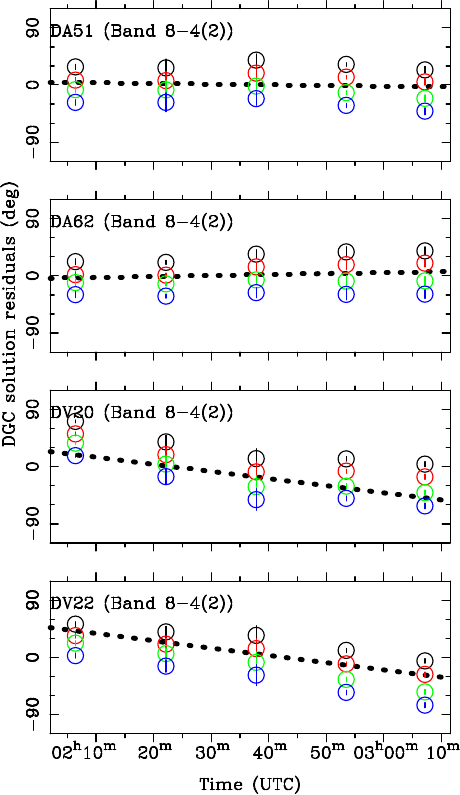}
\hspace{10mm}
\includegraphics[height=130mm]{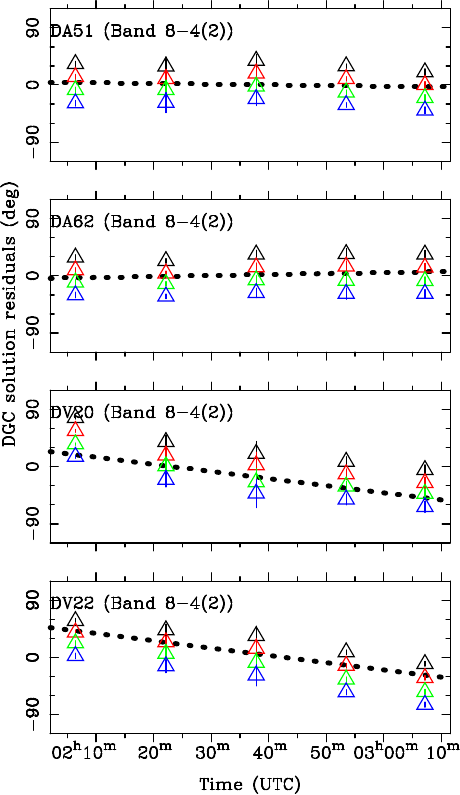}
\end{tabular}
\caption{
Same as 
%% FIGURE
Figure~\ref{fig:03} 
but at Band~8-4(2). 
}
\label{fig:04}
\end{center}
\end{figure}

\begin{figure}[htbp]
\begin{center}
\begin{tabular}{c}
\includegraphics[width=160mm]{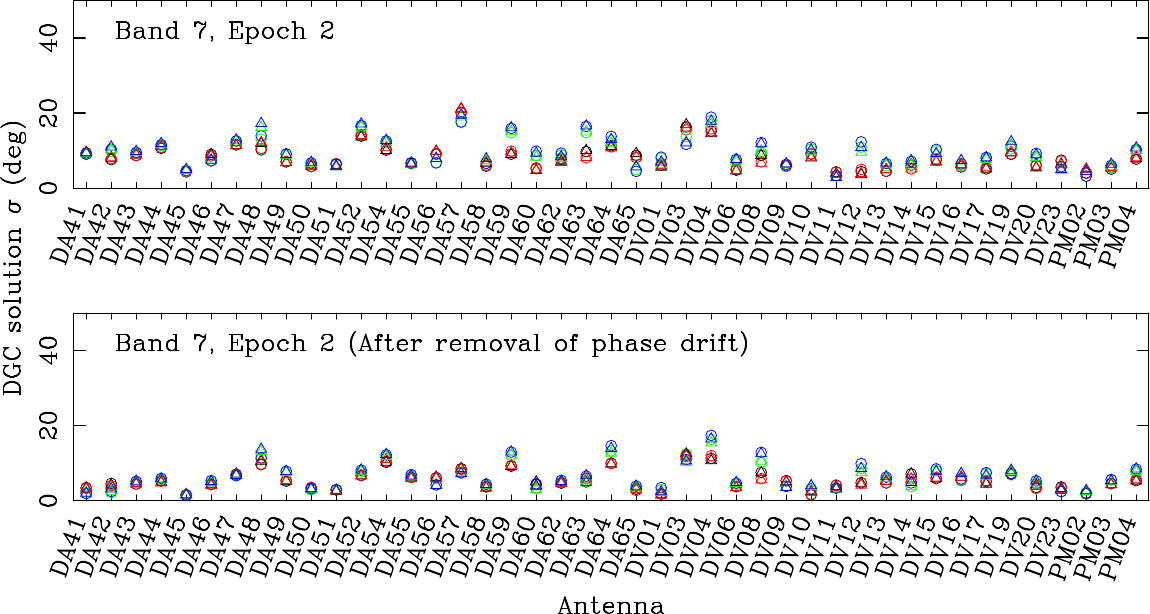}
\end{tabular}
\caption{
DGC solution RMS in Band~7-3(2). 
The top panel shows the RMS phase 
values in degrees 
for each SPW and each polarization pair. 
The bottom panel shows the RMS values of the residuals after removing 
the linear trend as described in 
%% SECTION
Section~\ref{sec:05-02} 
(examples of the linear trend are shown in 
%% FIGURE
Figure~\ref{fig:03} 
drawn with the dashed lines). 
The open circles and triangles represent $XX$ and $YY$ polarization pairs, respectively.  
The color differentiates four SPWs. 
}
\label{fig:05}
\end{center}
\end{figure}

\begin{figure}[htbp]
\begin{center}
\begin{tabular}{c}
\includegraphics[width=160mm]{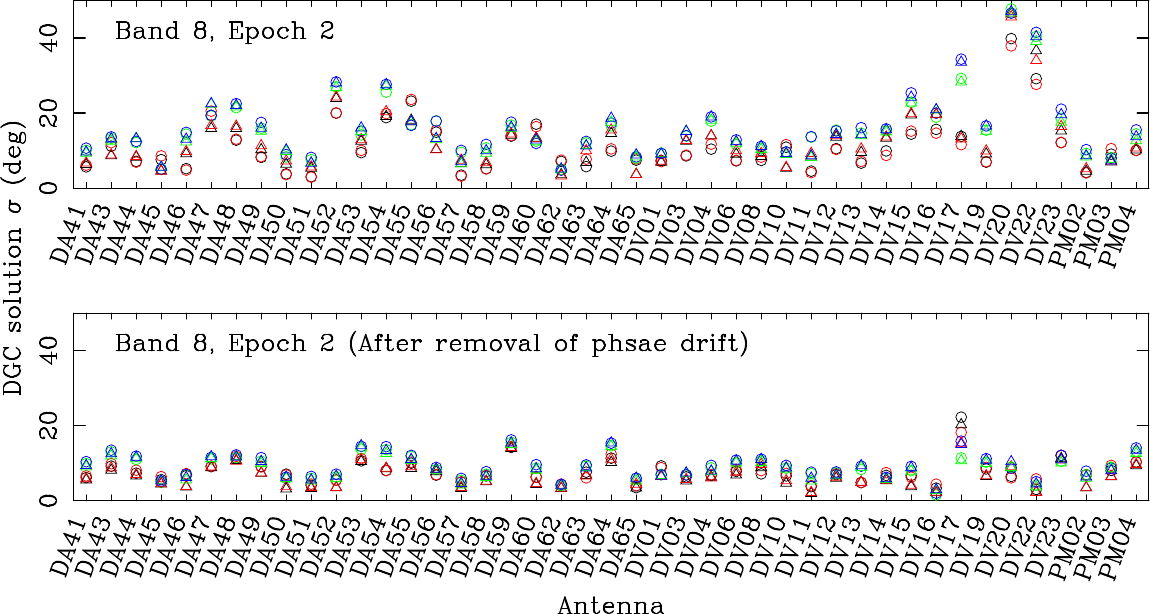}
\end{tabular}
\caption{
Same as 
%% FIGURE
Figure~\ref{fig:05} 
but in Band~8-4(2). 
}
\label{fig:06}
\end{center}
\end{figure}

%% FIGURE
Figures~\ref{fig:03} 
and 
%% FIGURE
\ref{fig:04} 
show the time variation of the antenna-based DGC solutions of the 
Band~7-3(2) and Band~8-4(2) experiments, respectively, for four specific 
antennas after subtracting the mean and then adding an arbitrary phase 
offset to each SPW for plotting purposes. 
The top two panels plot antennas with relatively small DGC solution RMS 
calculated for each of eight SPW/polarization pair combinations, while the 
bottom two panels plot other antennas with large DGC solution RMS. 
The DGC solution RMS is shown in the top panel of 
%% FIGURE
Figures~\ref{fig:05} 
and 
\ref{fig:06} 
at Band~7-3(2) and Band~8-4(2), respectively. 
We conducted a linear regression analysis for each antenna, including 
all the four SPWs and the two polarization pairs simultaneously, 
as displayed with dotted lines in each panel of 
%% FIGURE
Figures~\ref{fig:03} 
and 
%% FIGRE
\ref{fig:04}. 
In general, each antenna has a common trend in the time variation of the DGC 
solution for all the eight SPW/polarization pair combinations. Some of the antennas show 
a long-term linear trend 
(e.g. DV20 and DV22 in Band~8-4(2)) and/or a trend change 
(e.g. DA57 and DV04 in Band~7-3(2)). 
Generally speaking, such a long-term instability is hardly noticed in a single Band 
observation because the signal paths are completely the same for the corrected 
and correcting phases, i.e. if a long-term trend occurred, it would be present in both  
targets and a phase calibrator.  
On the other hand, in the case of B2B phase referencing, the signal path 
is partly different between the HF target and LF phase calibrator, as 
mentioned in 
%% SECTION
Section~\ref{sec:02}. 
One possible explanation of the long-term variation is because of 
a tiny time variation of the receiver physical temperature. 
ALMA front-end receiver cartridges are cooled  in the cryostat, and it is sometimes 
noticed that the temperature in a receiver can fluctuate over a time interval of hours.  
Such temperature time variations may be different at each receiver cartridge, so that the 
difference between the two frequencies would not be perfectly canceled out by 
differing the phases. 

Such a long-term instability can plausibly affect 
not only the DGC source but also the target and the phase 
calibrator in the same way, so that this linear trend can be corrected by 
an interpolation of the obtained 
DGC solutions. 
The bottom panels of 
%% FIGURE
Figures~\ref{fig:05} 
and 
%% FIGURE
\ref{fig:06} 
show 
the DGC solution RMS after subtracting the linear trend from the DGC solutions, 
which indicate an improvement of the order $10^{\circ}$ compared to without any 
linear correction. 
It is expected that, in our data reduction, 
the additional correction of the DGC solutions using 
the linear trend interpolation is useful to further stabilize the corrected phases 
when using B2B phase referencing. 

%%%%%%%%%%%%%
\subsection{
  Target source images
}\label{sec:05-03}
%%%%%%%%%%%%%

\begin{figure}[htbp]
\begin{center}
\begin{tabular}{c}
\includegraphics[width=150mm]{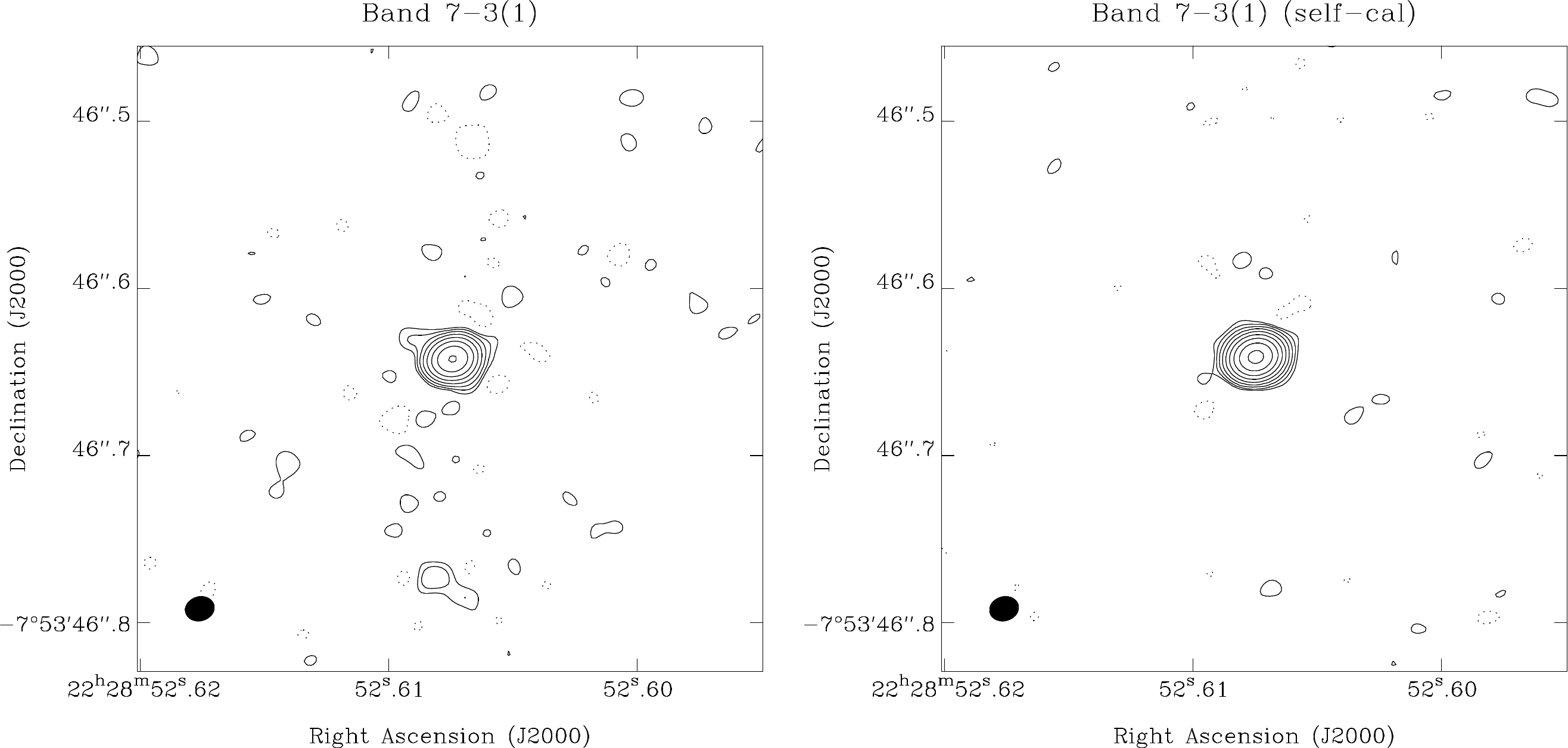}
\\
\includegraphics[width=150mm]{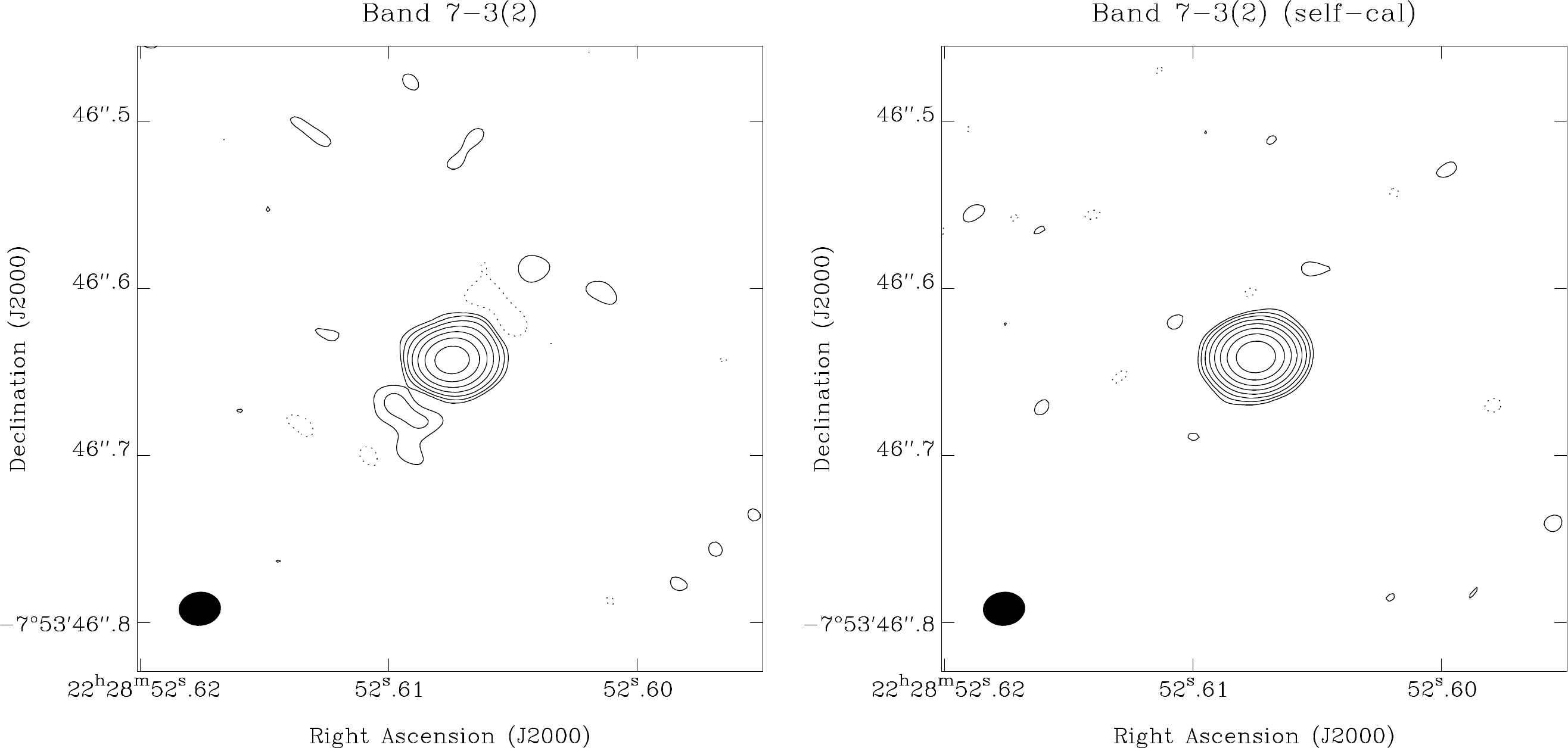}
\end{tabular}
\caption{
CLEAN images of the HF target J2228$-$0753 at Band~7-3. 
Top left: synthesis image of Band~7-3(1) with B2B phase referencing.
Top right: synthesis image of Band~7-3(1) with the phase self-calibration 
in addition to B2B phase referencing.
Bottom: synthesis images of Band~7-3(2) with B2B phase referencing (left) and 
with the phase self-calibration in addition to B2B phase referencing (right). 
The abscissa and ordinate are R.A. and decl., respectively. The contours are drawn 
for a factor of $-3$ (dash line), 3, 6, 12, 24, 48, 32, 64, and 128 (solid lines) multiplied by the 
rms noise of the images. 
The synthesized beam as listed in 
%% TABLE
Table~\ref{tbl:04} 
is shown in the lower left corner of each panel. The image peak values and image rms noises are also 
listed in 
%% TABLE
Table~\ref{tbl:04}. 
}
\label{fig:07}
\end{center}
\end{figure}

\begin{figure}[htbp]
\begin{center}
\begin{tabular}{c}
\includegraphics[width=150mm]{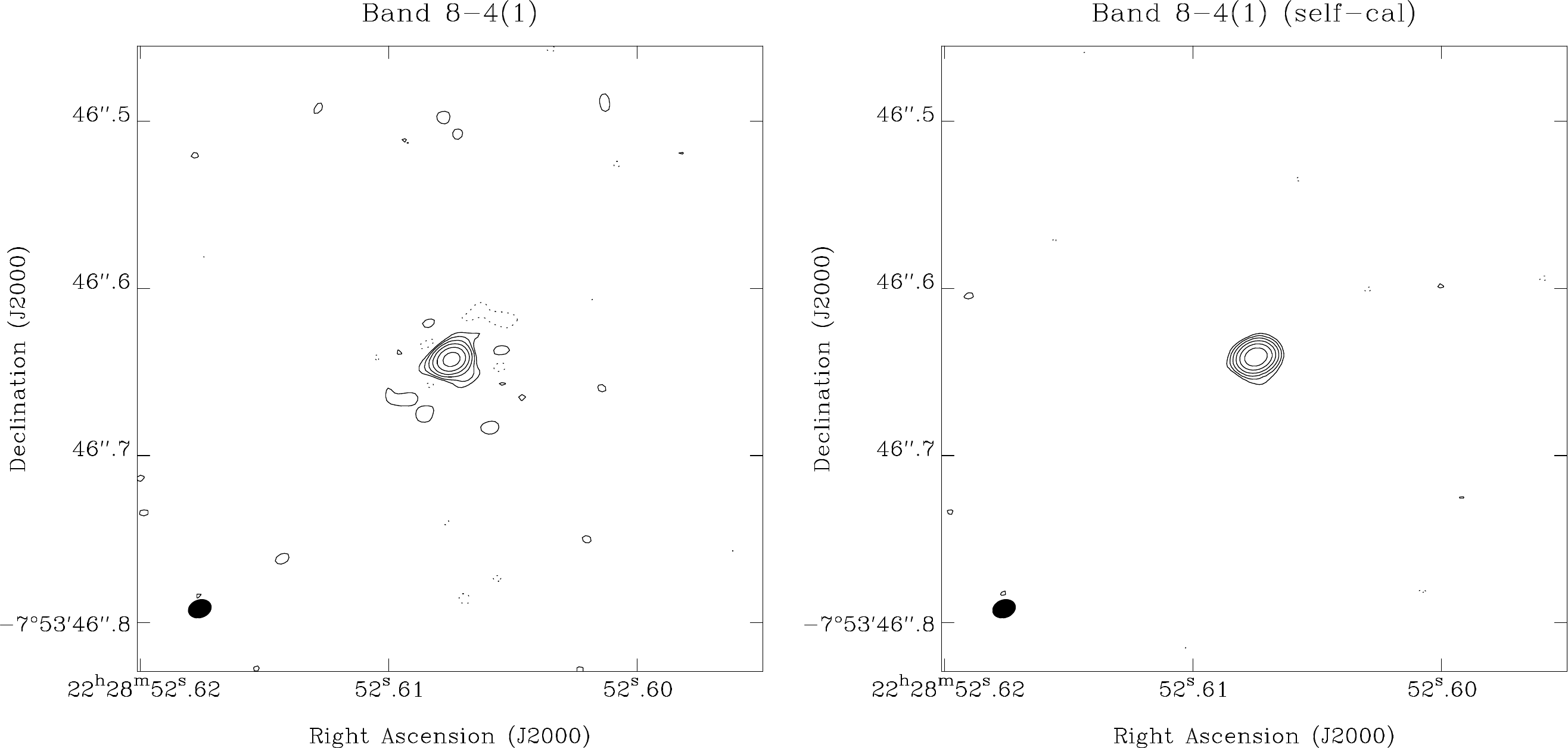}
\\
\includegraphics[width=150mm]{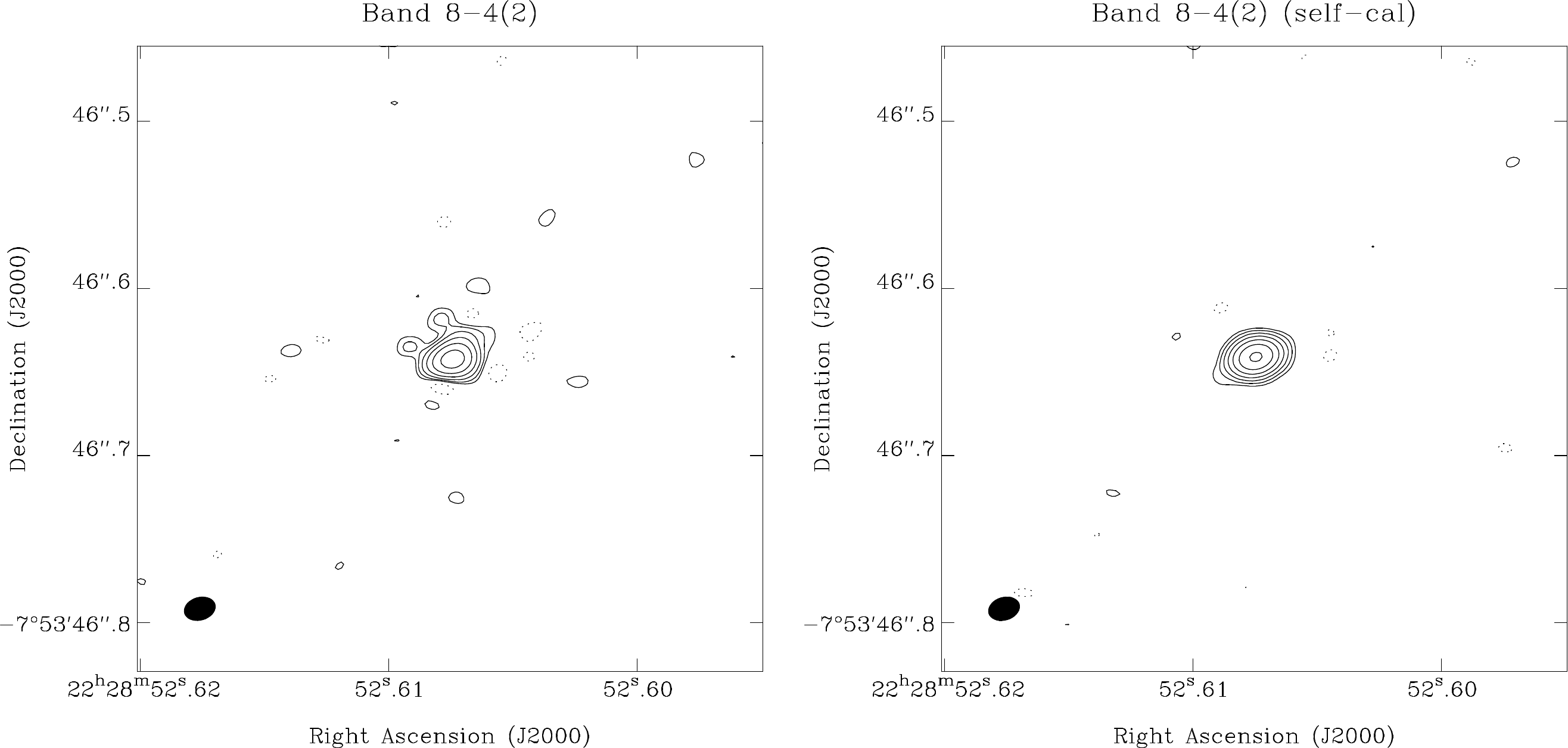}
\end{tabular}
\caption{
Same as 
%% FIGURE
Figure~\ref{fig:07}, 
but at Band~8-4.
}
\label{fig:08}
\end{center}
\end{figure}

The left panels of 
%% FIGURE
Figures~\ref{fig:07} 
and 
%% FIGURE
\ref{fig:08} 
show synthesized images of the target J2228$-$0753 at 289~GHz in Band~7-3 
and at 405~GHz in Band~8-4, respectively, after B2B phase referencing. 
The highest angular resolution of $14 \times 11$~mas was achieved at Band~8-4(1) 
with a maximum projected baseline length of $\sim 13$~km. 
The measured peak flux densities and image RMS noises are listed in 
%% TABLE
Table~\ref{tbl:04}. 
To verify the image quality using B2B phase referencing, 
phase self-calibration 
was additionally conducted after B2B phase referencing. The right panels of 
%% FIGURE
Figures~\ref{fig:07} 
and  
%% FIGURE
\ref{fig:08} 
show the images with the additional phase self-calibration. The peak flux densities 
of the self-calibrated images are consistent within 5\% between the two epochs 
in both Bands~7 and 8. Since a good amplitude repeatability is achieved in the 
synthesis images by reducing the phase errors as much as possible, the phase 
self-calibrated results can be regarded as a reference for the further analysis. 

One of the important indicators to evaluate the image quality is the image coherence, 
which is the ratio of the image peak flux density compared to the true value. 
We assume that the phase self-calibrated image represents the true value for 
each experiment. The obtained image coherence is listed in 
%% TABLE
Table~\ref{tbl:05}. 
The image coherence is 
larger than 90\% and 80\% at 289 and 405~GHz, respectively, 
with a close phase calibrator, 
so that in our tests B2B phase referencing with a close phase calibrator 
and 20~s switching cycle time proves to be an effective phase correction 
scheme for both our Band~7-3 and 8-4 experiments. 

We also investigated the target peak position in the synthesized images 
before self-calibration (right panels of 
%% FIGURE
Figures~\ref{fig:07} 
and 
%% FIGURE
\ref{fig:08}) 
using a two-dimensional Gaussian fitting in the image plane. 
The measured positions are listed in 
%% TABLE
Table~\ref{tbl:06}. 
The averaged position of J2228$-$0753 for the two epochs and two Bands is 
($\alpha_{\mathrm{J2000}}$,~$\delta_{\mathrm{J2000}}$)=($22^{\mathrm{h}}28^{\mathrm{m}}52^{\mathrm{s}}.607590$, $-7^{\circ}53'46''.64238$) 
with a $1\sigma$ error of (0.24,~0.40)~mas.
The ALMA calibrator source catalogue provides the position of the target as 
($\alpha_{\mathrm{J2000}}$,~$\delta_{\mathrm{J2000}}$)=($22^{\mathrm{h}}28^{\mathrm{m}}52^{\mathrm{s}}.60764$, $-7^{\circ}53'46''.6414$) 
with a $1\sigma$ error of (0.6,~1.2)~mas, so that 
the measured position of J2228$-$0753 is consistent with the position in the ALMA calibrator 
catalogue within the quoted 
uncertainties.
In centimeter-wave observations, 
J2228$-$0753's position measured with VLBI is 
($\alpha_{\mathrm{J2000}}$,~$\delta_{\mathrm{J2000}}$)=($22^{\mathrm{h}}28^{\mathrm{m}}52^{\mathrm{s}}.607568$, $-7^{\circ}53'46''.64215$) 
with a $1\sigma$ 
error of (0.10,~0.21)~mas from the 
Very Long Baseline Array (VLBA) 
calibrator list\footnote{
http://www.vlba.nrao.edu/astro/calib/}, 
thus, the obtained position is also consistent within 0.4 mas, less than a 20th of the Band~8-4 
synthesized beam.
As indicated by equation~(4) in 
%% REFERENCE
\cite{Asaki2020a}, 
a baseline vector error causes a systematic phase error after B2B phase referencing, 
and thus can cause an image distortion or an apparent positional shift of the target in the sky. 
The positional shift of the target $\Delta \mbox{\boldmath $\theta$}$ (rad) 
due to a baseline error vector $\Delta \mbox{\boldmath $\rho$}$ 
(in meters) is expressed by 
$\mbox{\boldmath $\rho$} \cdot \Delta \mbox{\boldmath $\theta$}$ $=$ 
$\Delta \mbox{\boldmath $\rho$} \cdot (\mbox{\boldmath $s$}_{\mathrm{t}} - \mbox{\boldmath $s$}_{\mathrm{c}})$, 
where 
$\mbox{\boldmath $\rho$}$ is a baseline vector and 
$\mbox{\boldmath $s$}_{\mathrm{t}}$ and $\mbox{\boldmath $s$}_{\mathrm{c}}$ are unit vectors 
from an observer to the target and phase calibrator in the sky, respectively.  
According to 
%% REFERENCE
\cite{Hunter2016}, 
$\Delta \rho$ is 2.5~mm for the longest baseline. 
For the target--phase calibrator pair of J2228$-$0753 and J2229$-$0832 with 
$\mid \mbox{\boldmath $s$}_{\mathrm{t}} - \mbox{\boldmath $s$}_{\mathrm{c}}\mid$ 
of $0^{\circ}.7$, 
%% 0.01187 rad
$\Delta \theta$ is equivalent to $\sim 0.4$~mas at most for a 16~km baseline, 
so that the obtained target position has a reasonable accuracy compared to 
the ALMA calibrator source catalogue and the VLBA calibrator list.

%%%%%%%%%%%%%
\section{
  Discussions
}\label{sec:06}
%%%%%%%%%%%%%
%%%%%%%%%%%%%%%%%%%%%%%
\subsection{
  Spatial Structure Function and Image Coherence 
  of the DGC Source
}\label{sec:06-01}
%%%%%%%%%%%%%%%%%%%%%%%

In the course of the data reduction, we obtained phase-corrected HF visibility data of the 
DGC source, J2253+1608. 
In a sense, DGC blocks are regarded as an ideal case of B2B phase referencing with the separation 
angle of $0^{\circ}$. 
In the case of these experiments, we expect that the phase correction of the target J2228$-$0753 
by the phase calibrator J2229$-$0832 will also be close to ideal, considering that the separation angle 
is only $0^{\circ}.7$. 
Thus, the HF DGC visibility data can be used as a proxy for 
the possible calibration that could be 
achieved for our target--phase calibrator pair.

\begin{figure}[htbp]
\begin{center}
\begin{tabular}{c}
\includegraphics[width=60mm]{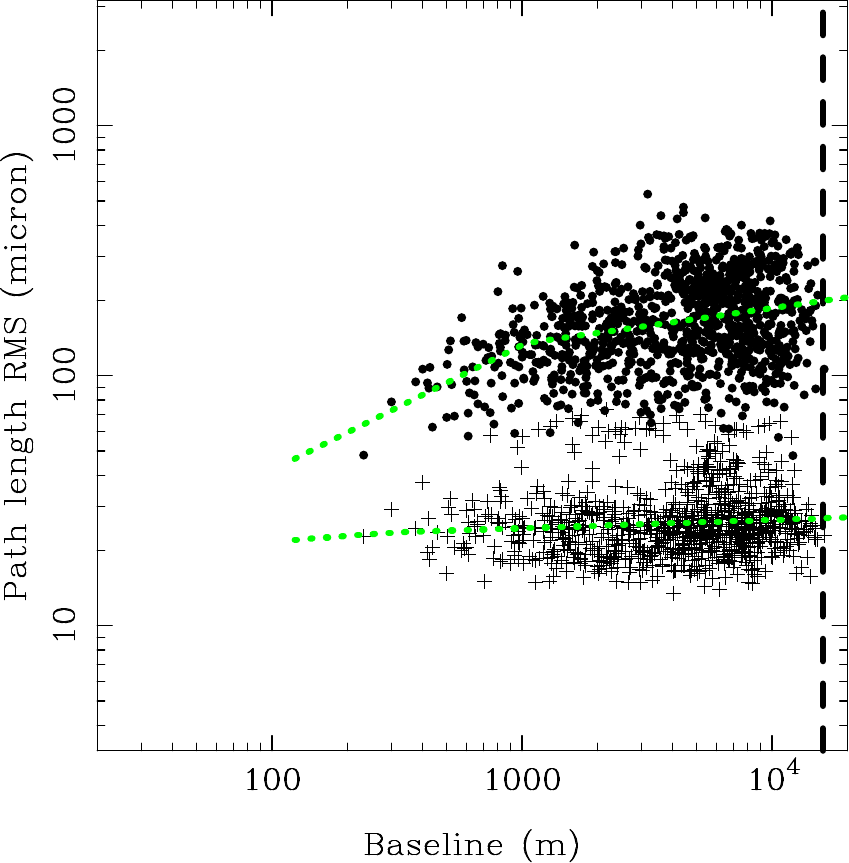}
\hspace{10mm}
\includegraphics[width=60mm]{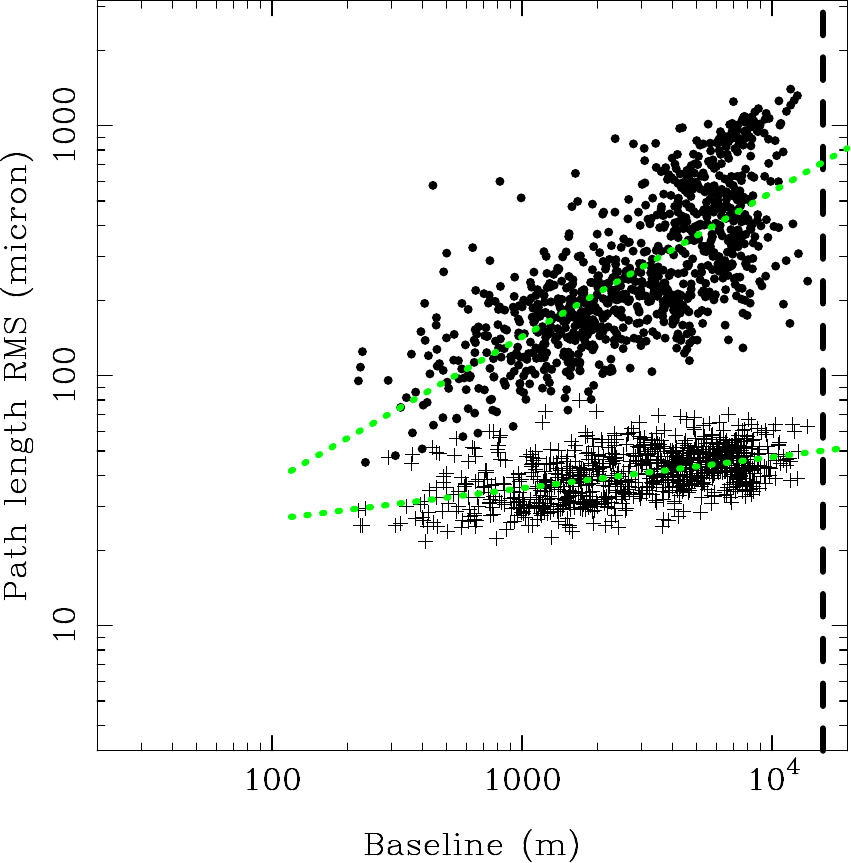}
\end{tabular}
\caption{
SSF of the Band~7-3 experiments. 
Left: SSF of the DGC source J2253$+$1608 in Band~3 (filled circles) and 
Band~7 (plus signs) at Band~7-3(1). 
Right: same as the left panel, but at Band~7-3(2). 
Note that the Band~3 phases were corrected with the WVR 
phase correction, while the Band~7 phases were corrected with the WVR phase 
correction and B2B phase referencing. The dotted lines are least-squares fitting 
results with two power-law components as mentioned in 
%% SECTION
Section~\ref{sec:06-01}. 
The vertical dashed lines indicate 16~km baseline length.
}
\label{fig:09}
\end{center}
\end{figure}

\begin{figure}[htbp]
\begin{center}
\begin{tabular}{c}
\includegraphics[width=60mm]{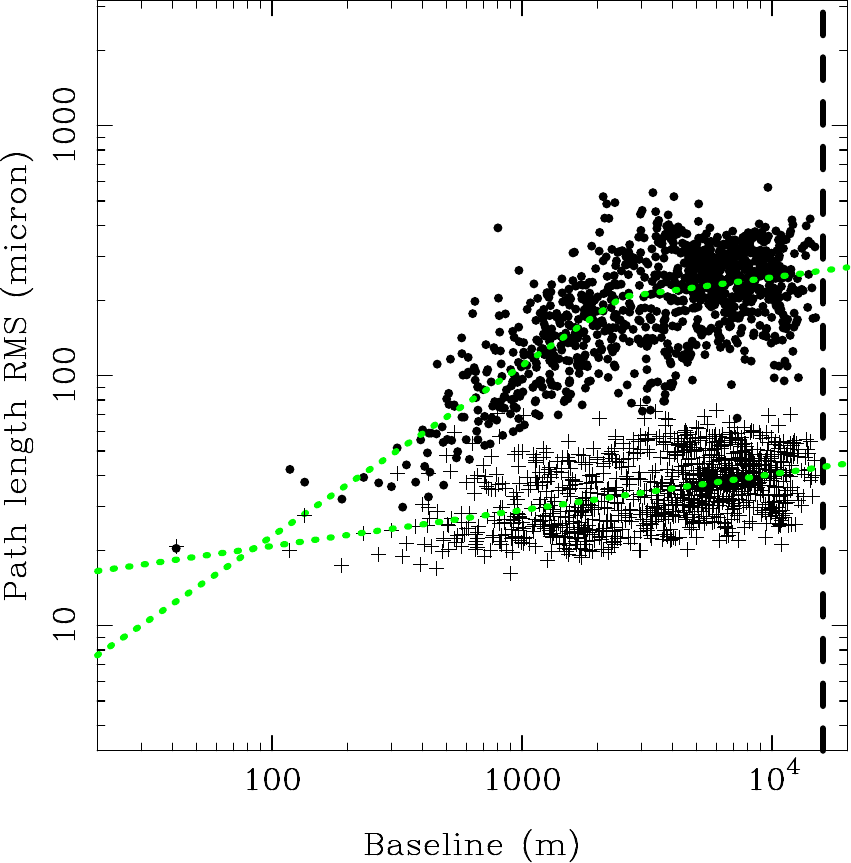}
\hspace{10mm}
\includegraphics[width=60mm]{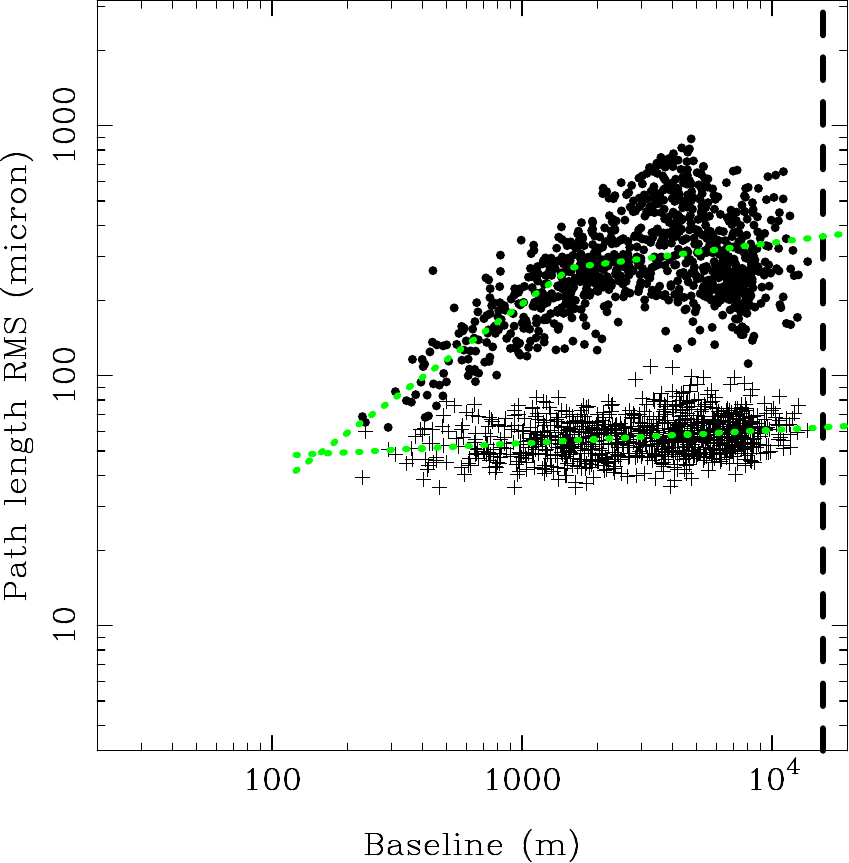}
\end{tabular}
\caption{
SSF of the Band~8-4 experiments. The designations are 
the same as in  
%% FIGURE
Figure~\ref{fig:09}. 
Left: SSF of the DGC source J2253$+$1608 in Band~4 (filled circles) 
and Band~8 (plus signs) at Band~8-4(1)
Right: same as the left panel, but at Band~8-4(2).  
}
\label{fig:10}
\end{center}
\end{figure}

In order to discuss the characteristics of the interferometer phase fluctuations, 
we introduce the spatial structure function (SSF) $D_{\mathrm{\Phi}}$  
which is a dispersion of interferometer phase as a function of 
a baseline length $\rho$ within the time interval of the whole observation duration 
%% REFERENCE
\citep{TMS2001}. 
The obtained square root of the SSF ($\sqrt{D_{\mathrm{\Phi}}(\rho)}$) 
as a function 
of the path length fluctuation is  shown in  
%% FIGURE
Figures~\ref{fig:09} and 
%% FIGURE
\ref{fig:10} 
for Band~7-3 and Band~8-4, respectively, using the LF DGC scans (filled circles) 
corrected with the WVR phase correction and 
HF DGC scans corrected the WVR phase correction and 
B2B phase referencing  (plus signs). 
%% REFERENCE
In general, the SSFs at the ALMA site 
show a turnover at baselines around 1~km, at which the power-law 
becomes shallower at longer baselines 
%% REFERENCE
\citep{Matsushita2017}. 
In order to evaluate the phase stability of ALMA's maximum baseline size (16~km), 
we fitted $\sqrt{D_{\mathrm{\Phi}}(\rho)}$ with a power-law 
function consisting of two components below and above the aforementioned 
turnover baseline.  
The fitted power-law functions are drawn with the dotted lines in 
%% FIGURE
Figures~\ref{fig:09} 
and 
%% FIGURE
\ref{fig:10}. 
The square roots 
of the SSFs for the shorter baseline range have the power of 0.4--0.7 
and the turnover baseline of 1--3~km before B2B phase referencing (the filled circles in 
%% FIGURE
Figures~\ref{fig:09} 
and 
%% FGIRE
\ref{fig:10}), 
roughly consistent with the values reported in 
%% REFERENCE
\cite{Matsushita2017}. 

After B2B phase referencing, the SSFs of the HF DGC scans 
decrease and are much less dependent on the baseline (the cross marks in 
%% FIGURE
Figures~\ref{fig:09} 
and 
%% FIGURE
\ref{fig:10}). 
The evaluated 16~km baseline phase RMS after B2B phase referencing are listed in 
%% TABLE
Table~\ref{tbl:05}. 
If the corrected DGC phases are a proxy for the correction achievable for the target, 
then we would expect that the target phase RMS would be 
reduced to the same level.
We imaged the HF DGC source to obtain the image coherence 
as listed in 
%% TABLE
Table~\ref{tbl:05}. 
Phase correction at 289~GHz  achieved 
a high image coherence of 99\% and 97\% for the Band~7-3(1) and 7-3(2) experiments, 
respectively. 
At 405~GHz, the DGC source image coherence is 95\% and 88\% for 
Bands~8-4(1) and 8-4(2), respectively.

In the Band~7-3 experiments, the path length fluctuations with a 
16~km baseline are 27 and 57~$\mu$m~rms (corresponding to 
the phase RMS of 0.16 and 0.35~radians at 289~GHz) 
at Band~7-3(1) and Band~7-3(2), respectively. 
These values correspond to the coherence factor of 99\% and 
94\%, consistent with the image coherence of the DGC source. 
In the case of Band~8-4, the path length fluctuations 
with a 16~km baseline are 42 and 62~$\mu$m~rms 
(corresponding to the phase rms of 0.35 and 
0.53~rad at 405~GHz), 
so that the coherence factor is 94\% and 87\% 
at Band~8-4(1) and Band~8-4(2), 
respectively, which is also consistent with the image coherence. 
Indeed, this is as one would expect given that the residual phases 
after phase correction are all that should remain for the visibility data 
and thus remain when creating the images. 
The SSFs and coherence factors of the 
phase-corrected HF DGC source on 
a 16~km baseline are also listed in 
%% TABLE
Table~\ref{tbl:05}.  

If we require an image coherence greater than 70\%, 
the phase stability should be higher than 0.84~rad. 
This implies that the path length fluctuations should be smaller than 
138 and 99~$\mu$m~RMS at the frequencies of 289 and 405~GHz, 
respectively.  
The Band~7-3(1) experiment shows that the SSF of the HF DGC source 
decreases to 27~$\mu$m~RMS in path length after B2B phase referencing. 
This indicates to us that if such a low path length is achievable, in similar 
observing conditions 
B2B phase referencing using ALMA long baselines at Band~9 (650~GHz) 
and Band~10 (950~GHz) could have achieved image coherence values of 
93\% and 87\%, respectively, if using a phase 
calibrator with a small separation angle of $<1^{\circ}$.

%%%%%%%%%%%%%%%%%%%%%%%
\subsection{
  Dependency of the image coherence on the switching cycle time
}\label{sec:06-02}
%%%%%%%%%%%%%%%%%%%%%%%

The residual atmospheric phase noise after phase referencing is 
$\sqrt{D_{\mathrm{\Phi}}(d + v_{\mathrm{w}}t_{\mathrm{swt}} /2)}$,
where 
$d + v_{\mathrm{w}}t_{\mathrm{swt}} /2$ is an equivalent baseline length 
of the phase SSF after phase referencing, 
$v_{\mathrm{w}}$ is the velocity of the atmosphere at the height of 
the turbulent layer, and $d$ is the geometrical distance between the lines of 
sight to the target and the 
phase calibrator at the altitude 
of the turbulent layer 
%% REFERENCE
\citep{Holdaway2004b}. 
Assuming that the altitude of the turbulent layer is 500~m and 
$v_{\mathrm{w}}$ is 6~m~s$^{-1}$ at the ALMA site
%% REFERENCE
\citep{Robson2001}, 
as well as a horizontal separation angle between the target and calibrator of $1^{\circ}$ 
at an elevation angle of $45^{\circ}$, 
then an equivalent baseline length is 
$12 + 3 t_{\mathrm{swt}}$, 
such that the dominant factor is the switching cycle time. 
With a shorter switching cycle time, it is clear that the residual phase is much lower given 
that the equivalent baseline length is obviously shorter. Furthermore, the SSF in the baseline 
regime $<$1--2~km increases with a power of $\sim 0.6$
%% REFERENCE
\citep{Matsushita2017}; 
thus, an increase in 
$t_{\mathrm{swt}}$ could noticeably increase the residual atmospheric phase noise 
after the phase correction. 

The baseline vector error affects the visibility phase after phase referencing as  
$\Delta \mbox{\boldmath $\rho$} \cdot (\mbox{\boldmath $s$}_{\mathrm{t}} - \mbox{\boldmath $s$}_{\mathrm{c}})$ 
(see 
%% SECTION
Section~\ref{sec:05-03}), 
so that the image quality is also affected by the separation angle. 
In the present study, we arranged the fixed target--phase calibrator pair with a 
small separation angle ($0^{\circ}.7$), which might not happen often: 
%% REFERENCE
\cite{Asaki2020a} 
report that the expected mean separation angle to find a suitable B2B phase referencing 
calibrator in Bands~3 to 7, 
to correct a science target observed in Bands~7 to 10, is $3^{\circ}$--$5^{\circ}$.  
Crucially, it is important that the phase calibrator is close on the sky 
as is explored further in 
%% REFERENCE
\cite{Maud2020}. 

Since we adopted the fixed 
switching cycle time of 20~s in the SBs, 
we here demonstrate a 60~s switching cycle time by flagging the last two scans 
in every three continuous phase calibrator scans in the early stage of the data 
reduction.
The Band~8-4 image coherence of the target was degraded to 84\% 
for Band~8-4(1) and 68\% for Band~8-4(2) with the 
60~s switching cycle time, while the Band~7-3 image coherence values were 
maintained to $\sim $90\%. 
The image coherence of the target and DGC source with the 60~s switching cycle time 
are listed in 
%% TABLE
Table~\ref{tbl:05}. 

The SSF and corresponding coherence factor at 16~km using the DGC visibility data 
are also listed in 
%% TABLE
Table~\ref{tbl:05}. 
The phase RMS at 16~km baseline and using a 60~s switching cycle time 
are 0.25 and 0.36~rad for Band~7-3(1) and Band~7-3(2), respectively, and 
0.55 and 0.85~rad for Band~8-4(1) and Band~8-4(2) respectively. The values are 
more consistent with the lower values of the previously mentioned 
2-minute phase stability (listed in 
%% TABLE
Table~\ref{tbl:03}). 
This suggests that the 2-minute phase stability is a convenient check 
parameter that can be used to roughly estimate the image quality of 
a long-baseline observation, provided that a 60~s switching cycle time and 
close phase calibrator are used.
As mentioned in 
%%SECTION
Section~\ref{sec:05-01}, 
the Band~8-4(2) experiment has a rather worse phase stability compared to the 
other experiments: 
the 2-minute phase stability reaches up to 1.62 rad during the middle of 
the Band 8-4(2) experiments,  
while the other experiments have the 2-minute phase stability below 1~rad. 
Note that only the 
Band~8-4(2) experiment shows an additional 14\% degradation in the image 
coherence compared to that with the 20~s switching cycle time, 
while the other experiments have only 4\%--5\% 
degradations. Thus, 
it is recommended for ALMA long-baseline observations that  
the go/no-go phase stability criterion ($<$ 1~radian) should be strictly kept 
if using $\sim 60$~s switching cycle time 
(e.g. the 72~s switching cycle time has been used for long-baseline user observations 
since ALMA opened the long baseline capability in Cycle~4), 
while for the 20~s switching cycle time we can relax the requirement to 
1--2~rad. In HF-LBC-2017, the B2B phase referencing image 
performance dependency on the switching cycle time was investigated in another 
series of experiments, which are detailed in 
%% REFERENCE
\cite{Maud2020}. 

The image coherence degradation as a function of the switching cycle time can be also 
verified using a temporal structure function (TSF) of the phase calibrator, which is defined by 
$\overline{D}_{\mathrm{\Phi}}(t_{\mathrm{swt}}, \rho) = \left < \left [ \Phi(t+t_{\mathrm{swt}}, \rho) - \Phi(t, \rho) \right ]^{2} \right >$, 
where $\Phi(t, \rho)$ is an interferometer phase of the phase calibrator with a baseline $\rho$ 
at time $t$. 
The TSF expresses 
the stability of the 
phase difference between two consecutive 
phase calibrator scans temporally separated by a switching cycle time. 
The TSF is computed for the phase calibrator at the LF and 
then scaled by the LF/HF frequency ratio to derive values for the HF. 
We evaluate the TSF for the baselines between the reference antenna and 
the outermost antennas (the upper quartile). 
For all the cases except for Band~8-4(2) with the 60~s switching cycle time, 
the TSFs are smaller than 1.5~rad. The TSF at Band~8-4(2) with the  60~s 
switching cycle time is $\sim 2.5$~rad when the significant image 
coherence degradation happens. 
The TSFs with the switching cycle times of 20~s and 60~s are listed in 
%% TABLE
Table~\ref{tbl:05}. 
Here, let us simplify the residual atmospheric phase error by ignoring the separation 
angle in this case, that is, $\sqrt{D_{\mathrm{\Phi}}(v_{\mathrm{w}}t_{\mathrm{swt}} /2)}$. 
Since $\sqrt{D_{\mathrm{\Phi}}(v_{\mathrm{w}}t_{\mathrm{swt}})}$
%square root of the SSF 
is a power-law function with a typical power of 0.6 for 
$v_{\mathrm{w}}t_{\mathrm{swt}} \leq 1$~km, 
%% REFERENCE
%\citep{Matsushita2017}, 
the residual atmospheric phase fluctuation of the target is 
expressed by 
$\sqrt{D_{\mathrm{\Phi}}(v_{\mathrm{w}}t_{\mathrm{swt}}/2)} 
\simeq 2^{-0.6} \times \sqrt{D_{\mathrm{\Phi}}(v_{\mathrm{w}}t_{\mathrm{swt}})}$. 
For $\rho \ll v_{\mathrm{w}}t_{\mathrm{swt}}$, the TSF can be translated to the SSF as 
$\overline{D}_{\mathrm{\Phi}}(t_{\mathrm{swt}}, \rho) \simeq 2 D_{\mathrm{\Phi}}(v_{\mathrm{w}}t_{\mathrm{swt}})$ 
%% REFERENCE
%\citep{Dravskikh1979}. 
\citep{Asaki1996}. 
As a result, 
the phase RMS of the target after phase referencing is 
$\sqrt{D_{\mathrm{\Phi}}(v_{\mathrm{w}}t_{\mathrm{swt}}/2)} 
\simeq 0.47 \times \sqrt{\overline{D}_{\mathrm{\Phi}}(t_{\mathrm{swt}}, \rho)}$. 
If the TSF is 1.80~rad in this simplified case, 
the residual atmospheric phase fluctuation is 0.84~rad, and thus 
we can obtain the image coherence of 70\%. In the case that the separation 
angle is too wide to be ignored (for example, wider than a few degrees), 
the requirement for the TSF is 
more severe, 
so that the criterion for the TSF would be $\sim 1$~rad 
with a given switching cycle time.

%%%%%%%%%%%%%%%%%%%%%%%
\subsection{
  Target On-source Time Efficiency
}\label{sec:06-03}
%%%%%%%%%%%%%%%%%%%%%%%

If a phase calibrator is very frequently observed at HF in order to mitigate 
the atmospheric phase fluctuations, overheads of such observations 
are considerably higher than standard ALMA observations, and the effective on-source 
time for the target can be a small fraction of the total. 
For example, 
in the cases of our second-epoch experiments, the target on-source time is 
12--13~min out of a total observation time of 1--1.5 hr, 
and thus the efficiency was only 15\%. 
One of the measures to improve the on-source time efficiency is to lengthen 
the switching cycle time. If the 2-minute phase stability is $< 1$~rad, 
a 60~s switching cycle time can be employed (see 
%% SECTION
Section~\ref{sec:06-02}), 
in which case the on-source efficiency will increase to 26\% 
for the same observing time as that in Band~7-3(2). 

If rather unstable weather conditions necessitate $\sim 20$~s switching cycle times,  
another strategy is to reduce the number of the DGC blocks.  
Our experimental SBs specifically included a larger number of DGC blocks in order to verify 
the DGC solution stability. As described in 
%% SECTION
Section~\ref{sec:05-02}, 
the solutions are stable with the long-term time variations that can be corrected. 
In the case that the DGC blocks are inserted only at the beginning and end of the SBs, 
the on-source time efficiency is improved to 25\% in Band~7-3(2). 
If we adopt the 60~s switching cycle time simultaneously, the on-source time efficiency 
is improved to 39\% in Band~7-3(2).

If a phase calibrator is bright enough, 
it would be possible to obtain a high enough S/N in a short scan 
at LF, so that the phase calibrator scan length can be reduced and, instead, 
the target scan length can be extended. In the case of narrow-bandwidth 
observations for millimeter/submillimeter molecular lines with a high velocity 
resolution, bandwidth switching to widen the bandwidth only for the phase calibrator 
frequency settings would be useful to raise the S/N. 
The combination of the bandwidth switching and B2B phase referencing techniques 
would provide the highest-possible S/N for the LF phase calibrator and 
thus could notably improve the target on-source time. Maintaining 
high image quality can be done at the expense of observing efficiency, and 
the amount of trade-off depends on the observing conditions and scientific goals.

%%%%%%%%%%%%%%%%%%%%%%%%%%%%%%%%%
\section{
  Summary
}\label{sec:07}
%%%%%%%%%%%%%%%%%%%%%%%%%%%%%%%%%

ALMA HF long-baseline experiments were conducted by adopting B2B phase 
referencing with a 20~s switching cycle time in order to demonstrate the ALMA 
high angular resolution image capability in Bands~7 and 8. 
In the series of experiments, 
we observed 
the target J2228--0753 paired with a phase calibrator J2229--0832 separated by 
$0^{\circ}.7$. 
We also observed J2253$+$1608 as the DGC source to remove the instrumental 
phase offset difference between the HF and LF 
Band receivers.

The DGC solutions are temporally stable for the receiver pairs of Band~7-3 and 
Band~8-4 for an hour with the phase RMS of $10^{\circ}$ after the correction of 
a linear trend. For Bands~7-3 and 8-4, the DGC solutions are so stable that 
the two DGC blocks are enough to insert at 
the beginning and end of an observation 
lasting $\sim 2$~hr. 

We achieved an 
angular resolution of $14 \times 11$~mas at 405~GHz 
for the target using the maximum array configuration with 16~km baselines. 
We obtained high image coherence above 98\% and 88\% at 289~GHz 
and 405~GHz, respectively. One of the experiments 
indicates a residual path length fluctuation of 27~$\mu$m RMS after B2B phase 
referencing, almost independent of baseline length, 
which would have met the requirements to achieve $>$70\% image coherence  
in Bands~9 and 10 using a phase calibrator with the separation angle of $< 1^{\circ}$.
The measured astrometric positions of the target are consistent with 
those in the ALMA calibrator source catalogue and the VLBA calibrator 
source list to within 0.4~mas. Because the separation angle between 
the target and phase calibrator was so small, that B2B 
phase referencing effectively reduced the phase error due to the baseline vector 
uncertainties, as well as the atmospheric phase fluctuations. 

We compared the image coherence achieved using 
the switching cycle times of 20 and 60~s, respectively. 
At 289~GHz, the image coherence did not significantly degrade, while at 405~GHz 
the image coherence decreased to 68$-$84\%. 
Observations that impose the condition that the 2-minute phase stability is 
$<1$ rad satisfy the requirement that the image coherence is $>$70\% 
when using 60~s switching cycle times. 
At higher frequencies, shorter switching cycle 
times will likely allow maintaining 
the coherence even if the 2-minute phase stability exceeds 1~rad.

\acknowledgments % equivalent to \section*{ACKNOWLEDGMENTS}
We used the following ALMA data: 
uid://A002/Xc5802b/X5bb3, uid://A002/Xc660ef/X8e0 
(J2228$-$0753 in Band~8 with B2B phase referencing), 
uid://A002/Xc59134/Xd47, uid://A002/Xc65717/X56f 
(J2228$-$0753 in Band~7 with B2B phase referencing).  
ALMA is a partnership of ESO (representing its member states), NSF (USA) and NINS (Japan), 
together with NRC (Canada), NSC and ASIAA (Taiwan), and KASI (Republic of Korea), in cooperation 
with the Republic of Chile. The Joint ALMA Observatory is operated by ESO, AUI/NRAO, and NAOJ. 
The authors thank to all the Joint ALMA Observatory staffs in Chile for performing the challenging 
HF-LBC-2017 successfully. 
L.~T.~M. was adopted as a JAO ALMA expert visitor during his stay. 
This work was supported by JSPS KAKENHI grant No. JP16K05306.

\software{
        CASA \citep{McMullin2007}
%       PGPLOT v5.2 \citep{PGPLOT}http://www.astro.caltech.edu/~tjp/pgplot/}
}
%
%Y.~A was financially supported by the East Asia ALMA 
%Regional Center of the National Astronomical Observatory of 
%Japan for staying in Chile to take part in the ALMA LBCs. 

% References
\bibliography{report} % bibliography data in report.bib
\bibliographystyle{aasjournal}

%\bibliographystyle{spiebib} % makes bibtex use spiebib.bst

%%%%%%%%%%%%%%%%%%%%%%%%%%%%%%%%%%%%%%%%%%%%
%%%%%%%%%%%%%%%%%%%%%%%%%%%%%%%%%%%%%%%%%%%%

\clearpage
\newpage

%%%%%%%%%%%%%%%%%%
%%%%%%%%%%%%%%%%%%

\clearpage
\newpage

%%%%%%%%%%%%%%%%%%
%%%%%%%%%%%%%%%%%%

\clearpage
\newpage

%%%%%%%%%%%%%%%%%%
%%%%%%%%%%%%%%%%%%

\clearpage
\newpage

%%%%%%%%%%%%%%%%%%
%%%%%%%%%%%%%%%%%%

\clearpage
\newpage

%%%%%%%%%%%%%%%%%%
%%%%%%%%%%%%%%%%%%

\clearpage
\newpage

%%%%%%%%%%%%%%%%%%
%%%%%%%%%%%%%%%%%%

\clearpage
\newpage

%%%%%%%%%%%%%%%%%%
%%%%%%%%%%%%%%%%%%

\clearpage
\newpage

%%%%%%%%%%%%%%%%%%%%%%%%%%%%%
%%%%%%%%%%%%%%%%%%

\clearpage
\newpage

%%%%%%%%%%%%%%%%%%
%%%%%%%%%%%%%%%%%%

\clearpage
\newpage

%%%%%%%%%%%%%%%%%%
%%%%%%%%%%%%%%%%%%

\clearpage
\newpage

%%%%%%%%%%%%%%%%%%
\begin{longrotatetable}
\begin{table}
\begin{center}
\caption{
  Experiment date and observing frequencies of the B2B phase referencing observation experiments
}\label{tbl:01}
\begin{threeparttable}
\begin{tabular}
{llcccccl}
\hline
\hline
Date 
& Experiment 
& Time Range
& Maximum 
& \multicolumn{1}{c}{Target} 
& \multicolumn{1}{c}{Calibrator}
& Target On-source
& Execution Block \\
%%%%

& Code
& (UT)
&  Unprojected
& LO1 Freq.
& LO1 Freq. 
& Duration
& \multicolumn{1}{c}{(uid://A002/)} \\
%%%%

&
&
&  Baseline Length (km)
& (GHz)
& (GHz)
& (s)
&  \\
%%%%
\hline
\multicolumn{8}{c}{First Epoch} \\
\hline
2017 Oct 9 
& Band~8-4(1)
& 03:09 $-$ 04:09     
& 16.20
& 405\tablenotemark{a}   
& 135\tablenotemark{b} 
& 341
&  Xc5802b/X5bb3 \\
%%%%
2017 Oct 10 
& Band~7-3(1)
& 03:02 $-$ 03:51   
& 16.20
& 289\tablenotemark{c}   
& 96\tablenotemark{d} 
& 355
&Xc59134/Xd47 \\
%%%%
\hline
\multicolumn{8}{c}{Second Epoch} \\
\hline
2017 Nov 2 
& Band~7-3(2)
& 00:57 $-$ 02:21  
&  13.89
& 289\tablenotemark{c}   
& 96\tablenotemark{d} 
& 411
& Xc65717/X56f \\
%%%%
2017 Nov 3 
& Band~8-4(2)
& 01:56 $-$ 03:08    
& 13.89
& 405\tablenotemark{a}   
& 135\tablenotemark{b} 
& 366
&  Xc660ef/X8e0 
\\
\hline
\end{tabular}
\tablenotetext{a}{The LO1 frequency is in Band~8.}
\tablenotetext{b}{The LO1 frequency is in Band~4.}
\tablenotetext{c}{The LO1 frequency is in Band~7.}
\tablenotetext{d}{The LO1 frequency is in Band~3.}
\end{threeparttable}
\end{center}
\end{table}
\end{longrotatetable}
%%%%%%%%%%%%%%%%%%

\clearpage
\newpage

%%%%%%%%%%%%%%%%%%
%\begin{landscape}
\begin{table}
\begin{center}
\caption{
  Observed sources of the B2B phase referencing observation experiments
}\label{tbl:02}
\begin{threeparttable}
\begin{tabular}
{llllcc}
\hline
\hline
Source Name
& Type
& R.A. 
& Decl. 
& Separation Angle 
& 
Flux Density
\\

& 
& (J2000) 
& (J2000)
& from the Target
& 
in Band~3
\\

&
&
&
& (deg)
& 
(Jy)
\\
\hline
%%%%
J2228$-$0753
& Target
& $22^{\mathrm{h}}28^{\mathrm{m}}52^{\mathrm{s}}.6076$
& $-7^{\circ}53'46.641''$
& $-$ 
& 
$0.18$\tablenotemark{a} 
\\
%%%%
J2229$-$0832
& Calibrator
& $22^{\mathrm{h}}29^{\mathrm{m}}40^{\mathrm{s}}.0834$
& $-8^{\circ}32'54.436''$
& $0.7$
& 
$2.59$\tablenotemark{b} 
\\
%%%%
J2253$+$1608 
& DGC source     
& $22^{\mathrm{h}}53^{\mathrm{m}}57^{\mathrm{s}}.7479$
& $+16^{\circ}08'53.561''$
& $24.8$ 
& 
$10.22$\tablenotemark{c} 
\\
\hline
\end{tabular}
\tablenotetext{a}{ALMA calibrator source catalogue at 91.5~GHz on 2016 October 16.}
\tablenotetext{b}{ALMA calibrator source catalogue at 91.5~GHz on 2017 May 5}
\tablenotetext{c}{ALMA calibrator source catalogue at 91.5~GHz on 2017 October 22}
\end{threeparttable}
\end{center}
\end{table}
%\end{landscape}
%%%%%%%%%%%%%%%%%%

\clearpage
\newpage

%%%%%%%%%%%%%%%%%%
%\begin{longrotatetable}
\begin{table}
\begin{center}
\caption{
  Weather conditions of the B2B phase referencing observation experiments.
}\label{tbl:03}
\begin{threeparttable}
\begin{tabular}
{lcccc}
\hline
\hline
Experiment 
& PWV
& 2-minute Phase Stability\tablenotemark{a}
& Wind Speed
& Wind Direction
\\
%%%%
Code
& (mm)
& (rad)\tablenotemark{b}
& (m~s$^{-1}$)
& (deg)
\\
%%%%
\hline
%\multicolumn{5}{c}{Band~7-3} \\
%\hline
%%%%
Band~7-3(1)
& 0.48
& 0.41--0.82      % 082, 0.42, 0.41
& 4.2
& 305
\\
%%%%
Band~7-3(2)
& 0.78
& 0.40--0.58    % 0.58, 0.40, 0.48, 0.44, 0.47, 0.40
& 5.3
& 318
\\
\hline
%\multicolumn{5}{c}{Band~8-4} \\
%\hline
%%%%
Band~8-4(1)
& 0.77
& 0.64--0.84    % 0.64, 0.74, 0.67, 0.84
& 8.1
& 314
\\
%%%%
Band~8-4(2)
& 0.59
& 0.67--1.62   % 0.96, 1.21, 1.62, 1.23, 0.67
& 4.8
& 320
\\
\hline
\end{tabular}
\tablenotetext{a}{Averaged phase RMS for 20\% longer baselines of the HF DGC scans 
in a single DGC block with  the time interval of 2 minutes. The minimum and maximum values are listed.}
\tablenotetext{b}{The representative frequency is 289 and 405~GHz for Bands~7-3 and 8-4, respectively.}
%\begin{tablenotes}\footnotesize
%\item[1] 
%Reference frequency is in ALMA RB Band~8.
%\item[8]
%\sout{Aborted due to the low elevation angle.}
%\end{tablenotes}
\end{threeparttable}
\end{center}
\end{table}
%\end{longrotatetable}
%%%%%%%%%%%%%%%%%%

\clearpage
\newpage

%%%%%%%%%%%%%%%%%%
%\begin{longrotatetable}
\begin{table}
\begin{center}
\caption{
  HF target J2228$-$0753 imaging results
}\label{tbl:04}
\begin{threeparttable}
\begin{tabular}
{lccc}
\hline
\hline
Experiment 
& Beam Size (mas)
& Peak Flux Density
& Peak Flux Density
\\
%%%%
Code
& (Position angle (deg))
& (Image Rms Noise)
& (Phase Self-alibrated)
\\
%%%%

& 
&  (mJy~beam$^{-1}$)
& (Image Rms Noise)
\\
%%%%

& 
&  
& (mJy~beam$^{-1}$)
\\
%%%%
\hline
%%%%
%%%%
%\multicolumn{5}{c}{Band~7-3} \\
%\hline
Band~7-3(1)
& $18 \times 15$ ($-74$)  
& 45.32 (0.11)
& 46.85 (0.05)
\\
%%%%
Band~7-3(2)
& $25 \times 20$ ($-85$) 
& 44.82 (0.15)
& 50.00 (0.07)
\\
\hline
%%%%
%%%%
%\multicolumn{5}{c}{Band~8-4} \\
%\hline
Band~8-4(1)
& $14 \times 11$ ($-70$)
%%%%%%%%%%%%%%& 30.96 (0.17)~mJy~beam$^{-1}$ 
& 34.63 (0.23)
& 37.90 (0.20)
\\
Band~8-4(2)
& $19 \times 14$ ($-73$)
& 31.37 (0.22)
& 38.84 (0.18)
\\
\hline
%%%%
%%%%
\end{tabular}
\end{threeparttable}
\end{center}
\end{table}
%\end{longrotatetable}
%%%%%%%%%%%%%%%%%%

\clearpage
\newpage

%%%%%%%%%%%%%%%%%%
\begin{longrotatetable}
\begin{table}
\begin{center}
\caption{
  Image coherence using 20~s and 60~s switching cycle times in B2B phase referencing
}\label{tbl:05}
\begin{threeparttable}
\begin{tabular}
{lccccccc}
\hline
\hline

& Switching 
& Target 
& \multicolumn{4}{c}{Phase-corrected HF DGC Scan} 
& Phase
\\
%%%%
\cline{4-7}
%%%%
Experiment 
& Cycle 
& Image 
& SSF\tablenotemark{b} (16~km)
& SSF\tablenotemark{b} (16~km) 
& Coherence Factor\tablenotemark{c} 
& Image 
& Calibrator
\\
%%%%
Code
& Time (s)
& Coherence\tablenotemark{a} (\%)
& ($\mu$m RMS)
& (rad)
& (16~km) (\%)
& Coherence\tablenotemark{a} (\%)
& TSF\tablenotemark{d} (rad)
\\
\hline
\multirow{2}{*}{Band~7-3(1)}
& 20
& 97
& 27
& 0.16 % 9deg
& 99
& 99
& 0.48
\\
%\cline{2-8}
%%%%

& 60
& 92 
& 40
& 0.25 % 14deg
& 97 
& 96 
& 0.93
\\
%%%%
\hline
\multirow{2}{*}{Band~7-3(2)}
& 20
& 93
& 57
& 0.35 % 20deg
& 94
& 97
& 0.57
\\
%\cline{2-8}
%%%%

& 60
& 89 
& 59
& 0.36 %20 deg
& 94 
& 94 
& 1.05
\\
\hline
%\multicolumn{3}{c}{Band~8-4} \\
%\hline
%%%%
\multirow{2}{*}{Band~8-4(1)}
& 20
& 87
& 42
& 0.36 % 20 deg
& 94
& 95
& 0.75
\\
%\cline{2-8}
%%%%

& 60
& 84 
& 65
& 0.55 % 32 deg
& 86 
& 91 
& 1.23
\\
%%%%
\hline
\multirow{2}{*}{Band~8-4(2)}
& 20
& 82
& 62
& 0.53 % 30 deg
& 87
& 88
& 1.47
\\
%\cline{2-8}
%%%%

& 60
& 68 
& 100
& 0.85 % 49 deg
& 70 
& 74 
& 2.52
\\
\hline
\end{tabular}
\tablenotetext{a}{Image coherence is the ratio of the image peak flux density compared to 
the true value. See Sections~\ref{sec:05-03}, \ref{sec:06-01}, and \ref{sec:06-02} in detail.}
\tablenotetext{b}{SSF: spatial structure function. 
See Sections~\ref{sec:06-01} and \ref{sec:06-02} in detail.}
\tablenotetext{c}{Coherence factor is calculated by $\exp{(-\sigma_{\Phi}^{2}/2)}$, 
where $\sigma_{\Phi}$ is the phase RMS with a 16~km baseline. 
See Sections~\ref{sec:01}, ~\ref{sec:06-01}, and \ref{sec:06-02} in detail.}
\tablenotetext{d}{TSF: temporal structure function. See Section~\ref{sec:06-02} in detail.}
\end{threeparttable}
\end{center}
\end{table}
\end{longrotatetable}
%%%%%%%%%%%%%%%%%%

\clearpage
\newpage

%%%%%%%%%%%%%%%%%%
%\begin{landscape}
\begin{table}
\begin{center}
\caption{
  Position measurements of the target J2228$-$0753 image peak
}\label{tbl:06}
\begin{threeparttable}
\begin{tabular}
{lll}
\hline
\hline
Experiment Code
& R.A. ($1\sigma$ Error)
& Decl. ($1\sigma$ Error)
\\

& (J2000) 
& (J2000)
\\
\hline
%\multicolumn{3}{c}{Band~7-3} \\
%\hline
%%%%
%2017 Oct 10
%& $22^{\mathrm{h}}28^{\mathrm{m}}52^{\mathrm{s}}.6076$
%& $-07^{\circ}53'46.643''$
%\\
Band~7-3(1)
&  $22^{\mathrm{h}}28^{\mathrm{m}}52^{\mathrm{s}}.607559$  (0.02 mas) 
& $-7^{\circ}53'46''.64224$  (0.02 mas)
\\
%%%%
%2017 Nov 2
%& $22^{\mathrm{h}}28^{\mathrm{m}}52^{\mathrm{s}}.6076$
%& $-07^{\circ}53'46.643''$
%\\
Band~7-3(2)
&  $22^{\mathrm{h}}28^{\mathrm{m}}52^{\mathrm{s}}.607570$  (0.04 mas) 
& $-7^{\circ}53'46''.64270$  (0.03 mas)
\\
\hline
%\multicolumn{3}{c}{Band~8-4} \\
%\hline
%%%%
%2017 Oct 9 
%& $22^{\mathrm{h}}28^{\mathrm{m}}52^{\mathrm{s}}.6076$
%& $-07^{\circ}53'46.643''$
%\\
Band~8-4(1)
&  $22^{\mathrm{h}}28^{\mathrm{m}}52^{\mathrm{s}}.607606$  (0.04 mas) 
& $-7^{\circ}53'46''.64248$  (0.03 mas)
\\
%%%%
%2017 Nov 3 
%& $22^{\mathrm{h}}28^{\mathrm{m}}52^{\mathrm{s}}.6076$
%& $-07^{\circ}53'46.643''$
%\\
Band~8-4(2)
&  $22^{\mathrm{h}}28^{\mathrm{m}}52^{\mathrm{s}}.607550$  (0.06 mas) 
& $-7^{\circ}53'46''.64215$  (0.04 mas)
\\
%%%%
\hline
\end{tabular}
\end{threeparttable}
\end{center}
\end{table}
%\end{landscape}
%%%%%%%%%%%%%%%%%%

\end{document}